\def\la{\mathrel{\mathchoice {\vcenter{\offinterlineskip\halign{\hfil
$\displaystyle##$\hfil\cr<\cr\sim\cr}}}
{\vcenter{\offinterlineskip\halign{\hfil$\textstyle##$\hfil\cr
<\cr\sim\cr}}}
{\vcenter{\offinterlineskip\halign{\hfil$\scriptstyle##$\hfil\cr
<\cr\sim\cr}}}
{\vcenter{\offinterlineskip\halign{\hfil$\scriptscriptstyle##$\hfil\cr
<\cr\sim\cr}}}}}

\def\ga{\mathrel{\mathchoice {\vcenter{\offinterlineskip\halign{\hfil
$\displaystyle##$\hfil\cr>\cr\sim\cr}}}
{\vcenter{\offinterlineskip\halign{\hfil$\textstyle##$\hfil\cr
>\cr\sim\cr}}}
{\vcenter{\offinterlineskip\halign{\hfil$\scriptstyle##$\hfil\cr
>\cr\sim\cr}}}
{\vcenter{\offinterlineskip\halign{\hfil$\scriptscriptstyle##$\hfil\cr
>\cr\sim\cr}}}}}

\def\p {$\pm$}

\def\kms {\hbox{${\rm km\, s}^{-1}$}} % km s-1 for LATEX
\def\cmsq  {$\hbox{{\rm cm}}^{-2}$}    %cm-2
\def\arcsec {\hbox{$^{\prime\prime}$}}

\def\percc {$\hbox{{\rm cm}}^{-3}$}    %cm-3
\def\MOLH {\hbox{${\rm H}_2$}}  %H2
\def\MOLN {\hbox{${\rm N}_2$}}  % N2
\def\AMM {\hbox{${\rm NH}_{3}$}} %NH3
\def\HCOP {\hbox{${\rm HCO}^+$}}      %HCO+
\def\HTHCOP {\hbox{${\rm H^{13}CO}^+$}}      %H13CO+
\def\HCEIOP {\hbox{${\rm HC^{18}O}^+$}}      %HC18O+
\def\HCSEOP {\hbox{${\rm HC^{17}O}^+$}}      %HC17O+
\def\DCOP {\hbox{${\rm DCO}^+$}}    %DCO+
\def\DTHCOP {\hbox{${\rm D^{13}CO}^+$}}    %D13CO+
\def\CEIO {\hbox{${\rm C}^{18}{\rm O}$}}   %C18O
\def\CSEO {\hbox{${\rm C}^{17}{\rm O}$}}   % C17O
\def\NTHP {\hbox{${\rm N}_2{\rm H}^+$}} % N2H+
\def\NTDP {\hbox{${\rm N}_2{\rm D}^+$}} % N2D+
\def\CYAC {\hbox{${\rm HC}_3{\rm N}$}}     %HC3N
\def\HNFINP {\hbox{${\rm HN}^{15}{\rm N}^+$}} % HN15N+
\documentstyle[11pt,aaspp4]{article}
\lefthead{Caselli et al.}
\righthead{The ionization degree in L~1544}

\begin{document}

\title{Molecular ions in L~1544. I. Kinematics}

\author{P. Caselli , C. M. Walmsley, \and A.Zucconi}
\affil{Osservatorio Astrofisico di Arcetri, Largo E. Fermi 5, I-50125
Firenze, Italy; caselli@arcetri.astro.it; walmsley@arcetri.astro.it;
zucconi@arcetri.astro.it}

\author{M. Tafalla}
\affil{Observatorio Astron\'omico Nacional (IGN), Campus Universitario,
E-28800 Alcal\'a de Henares (Madrid), Spain; tafalla@oan.es}

\author{L. Dore}
\affil{Dip. Chimica "Ciamincian", Universit\`{a} di Bologna,
Via Selmi 2, I-40126, Bologna, Italy; dore@ciam.unibo.it}

\and

\author{P. C. Myers}
\affil{Harvard--Smithsonian Center for Astrophysics, MS 42, 60
Garden Street, Cambridge, MA 02138, U.S.A.; pmyers@cfa.harvard.edu}

% Notice that each of these authors has alternate affiliations, which
% are identified by the \altaffilmark after each name.  The actual alternate
% affiliation information is typeset in footnotes at the bottom of the
% first page, and the text itself is specified in \altaffiltext commands.
% There is a separate \altaffiltext for each alternate affiliation
% indicated above.

%\altaffiltext{1}{Visiting Astronomer, Cerro Tololo Inter-American Observatory.
%CTIO is operated by AURA, Inc.\ under contract to the National Science
%Foundation.}
%\altaffiltext{2}{Society of Fellows, Harvard University.}
%\altaffiltext{3}{present address: Center for Astrophysics,
%    60 Garden Street, Cambridge, MA 02138}
%\altaffiltext{4}{Visiting Programmer, Space Telescope Science Institute}
%\altaffiltext{5}{Patron, Alonso's Bar and Grill}

% The abstract environment prints out the receipt and acceptance dates
% if they are relevant for the journal style.  For the aasms style, they
% will print out as horizontal rules for the editorial staff to type
% on, so long as the author does not include \received and \accepted
% commands.  This should not be done, since \received and \accepted dates
% are not known to the author.

\begin{abstract}
We have mapped the dense dark core L~1544 in \HTHCOP (1--0), \DCOP (2--1),
\DCOP (3--2), \NTHP (1--0), \NTHP (3--2), \NTDP (2--1), \NTDP (3--2),
\CEIO (1--0), and \CSEO (1--0) using the IRAM 30--m telescope.  We have
obtained supplementary observations of \HCEIOP (1--0), \HCSEOP (1--0), and
\DTHCOP (2--1). Many of the observed maps show a general correlation
with the distribution of dust continuum emission in contrast to
\CEIO (1--0) and \CSEO (1--0) which give clear evidence for depletion of
CO at positions close to the continuum peak.  In particular \NTDP  (2--1)
and (3--2) and to a lesser extent \NTHP (1--0)
appear to be excellent tracers of the
dust continuum.  Our \DCOP \ maps have the same general morphology as
the continuum while \HTHCOP (1--0) is more extended.
We find also that many apparently optically
thin spectral lines  such as \HCEIOP \ and \DTHCOP \ have double
or highly asymmetric profiles towards the dust continuum peak.
We have studied the velocity field in the high density nucleus of
L~1544 putting particular stress on tracers such as \NTHP \
and \NTDP \ which trace the dust emission and which we therefore
believe trace the gas with density of order of $10^5$ \percc .
We find that the tracers of high density gas 
(in particular \NTDP ) show a velocity gradient
along the minor axis of the L~1544 core and that there is evidence for
larger linewidths close to the dust emission peak. We interpret this
using the model of the L~1544 proposed by ~\cite{CB00} and by comparing
the observed velocities with those expected on the basis of their
model. The results show reasonable agreement between observations
and model in that the velocity gradient along the minor axis and 
the line broadening toward the center of L~1544 are  
predicted by the model.  This is evidence in favour of the idea that amipolar 
diffusion across field lines is one of the basic processes leading to 
gravitational collapse.  However, the double--peaked nature of the profiles is
reproduced only at the core center and if a ``hole'' in the molecular 
emission, due to depletion, is present.  Moreover, line widths are 
significantly 
narrower than observed and are better reproduced by the ~\cite{MZ01} model
which considers the quasistatic vertical contraction of a layer due to 
dissipation of its Alfv\'enic turbulence, indicating the importance of 
this process for cores in the verge of forming a star.

\end{abstract}

% The different journals have different requirements for keywords.  The
% keywords.apj file, found on aas.org in the pubs/aastex-misc directory,
% contains a list of keywords used with the ApJ and Letters.  These are
% usually assigned by the editor, but authors may include them in their
% manuscripts if they wish.

%\keywords{globular clusters,peanut clusters,bosons,bozos}

% That's it for the front matter.  On to the main body of the paper.
% We'll only put in tutorial remarks at the beginning of each section
% so you can see entire sections together.

% In the first two sections, you should notice the use of the LaTeX \cite
% command to identify citations.  The citations are tied to the
% reference list via symbolic KEYs.  We have chosen the first three
% characters of the first author's name plus the last two numeral of the
% year of publication.  The corresponding reference has a \bibitem
% command in the reference list below.
%
% Please see the AASTeX manual for a more complete discussion on how to make
% \cite-\bibitem work for you.

\section{Introduction}

 A considerable amount of observational work has gone  into the
 study of ``prestellar cores'' (sometimes called pre--protostellar
 cores) which are dense condensations within molecular clouds which
 are thought to be on the point of collapse (see e.g. ~\cite{AWB00}). 
 Such objects  are found observationally to be extremely cold
 ($<$ 15 K) and with column densities much higher than the surrounding
 molecular clouds.  They also occasionally show evidence for
 infall (~\cite{TMM98}) and an example of this is the
 high density core associated with L~1544.  One concludes 
 therefore that such objects are close to being in the 
 ``pivotal state'' from which gravitational collapse can
 commence.  Determining their density distribution and 
 kinematic properties is thus a matter of great interest. 
  
 The density distribution in ``starless cores'' has been delineated
 both by the observed dust emission
 and extinction as well as in molecular line maps.  The latter however
 appear in many cases to give an inaccurate picture of the true
 column density distribution due to the fact that at the temperatures
 of the pre--stellar cores, much of the molecular gas is depleted out
 onto dust grains.  The dust emission/extinction data 
 (e.g \cite{WMA99} (hereafter WMA),~\cite{BAP00} (hereafter BAP),
 ~\cite{ALL01})  show that starless cores
 have molecular hydrogen column densities ranging up to $10^{23}$
 \cmsq \ equivalent to  100 magnitudes of visual extinction. 
 There is evidence for a  column density profile  varying as
 a function of offset from the central position $R$ as $1/R$ 
 but with a flatter profile below a critical radius of roughly 3000
 AU.  This is reminiscent of the structures expected in hydrostatic
 equilibrium such as a ``Bonnor--Ebert sphere'' and indeed 
 recent NIR extinction data for B68 show impressive agreement with
 the distribution expected for such a sphere (~\cite{ALL01}). 
 However, it is clear that reality is more complicated since the
 contours of dust emission/absorption as a rule are far from
 being spherically symmetric. This suggests that magnetic fields
 play an important role in the dynamics of such systems. 

 Models for the likely magnetic field structures in cores 
 in quasi--hydrostatic equilibrium have been constructed by a
 variety of authors (see e.g ~\cite{SAL87},~\cite{CB00} 
 (hereafter CB00), ~\cite{CB01} and
 references therein). These  vary in detail but have in
 common the idea that ambipolar diffusion across magnetic field
 lines permits collapse on a time scale roughly an order of
 magnitude longer than the free--fall time. In this case,
 one expects a ``disk--like'' structure and for the particular case
 of L~1544, it has been shown (CB00)  that the observed dust column
 density distribution is in reasonable agreement with the model   
 column density distribution just prior to collapse.  In their 
 model, the observed ``disk'' is almost edge--on (inclined at an
 angle of $16^{\circ}$) and is contracting radially in the plane of
 the disk.  Recently, \cite{MZ01} (hereafter MZ) have studied a
flattened cloud in quasistatic vertical contraction triggered by 
Alfv\'enic turbulence dissipation.  As we will see, 
this model also reproduces some of the features observed towards
L~1544.

 Understanding the kinematics in cores such as L~1544 clearly
 requires observation of molecular lines. The results of
 ~\cite{CWT99} (see also ~\cite{CW01})
 show clearly however that CO and its isotopically
 substituted counterparts are depleted onto dust grain surfaces at
 densities above $10^5$ \percc \ in L~1544. Observations of other  
 cores (\cite{TMC01}) show a very similar picture. 
 However it is clear that certain species and, in particular
 \NTHP ,  have a spatial distribution which is much more similar
 to that of the dust than CO. The natural inference is that these
 species, or perhaps species to which they are chemically linked,
 are either undepleted or much less depleted than CO at high
 densities. In the particular case of \NTHP , it is known chemically
 to be linked to molecular nitrogen which is likely to be the
 main repository of nitrogen in the gas phase. \MOLN \ 
 is somewhat  more volatile than CO and hence it is plausible
 that there are situations where \MOLN \ is undepleted and CO
 is in solid form (see e.g. ~\cite{BL97}).  This has the consequence
 that the kinematics of dense gas in starless cores such as L~1544
 will be best delineated by species such as \NTHP \ and chemically
 related molecules (e.g. \NTDP ).  One of the objectives of the
 present study is to test current models of the contraction of
 L~1544 exploiting this fact.

 L~1544 consists of a dense core surrounded by a low density
 envelope which causes self--absortion of many transitions emitted
 by the dense gas.  This self--absorption is a considerable problem
 for studies aimed at determining the characteristics of the denser
 material. The \NTHP (1--0) transition shows evidence for 
 absorption (~\cite{WMW99}) by a foreground layer of density 
 roughly $10^4$ \percc \ which is red--shifted by 0.1 \kms . 
 This foreground absorption is a complication for those
 interested in the kinematics of the dense gas and there are
 therefore advantages to be gained in observing higher J
 transitions of \NTHP \ and analogous species for which
 it is likely that the foreground absorption will be either
 absent or very weak. 
 
  In the present article, we describe a study of L~1544 
 carried out in a variety of molecular transitions. 
 We have concentrated on the emission of molecular ions because 
 a main aim of the study
 has been to determine the ionization degree as a function of
 position within the dense nucleus of L~1544. This parameter
 essentially determines the ambipolar diffusion time scale
 and hence the timescale for star formation. A companion 
 paper to the present article (~\cite{CZWb}, hereafter Paper II)
 concerns itself with the determination
 of the ionization degree and the chemistry of the high density gas
 in L~1544. In this article, we present the data and discuss
 what conclusions can be drawn about the kinematics of the dense
 gas.   We conclude
 by making a comparison of our data with that expected on the
 basis of models of core evolution.

Observations are described in Section 2 and 
our principal observational results are discussed in Section 3. 
In Section 4, we
compare with the models of CB00 and MZ and discuss the outcomes. In 
 section 5, we summarize  our conclusions.

\section{Observations}

 The data described here were taken using the IRAM 30-m telescope in
 three sessions: January 1997, August 1997, November 1998, and July 2000.
 We in general observed using simultaneously the three facility
 receivers at 3, 2, and 1.3mm. 
Part of the \NTHP (1--0) data reported and analysed in this
paper have been already presented in ~\cite{TMM98}.
We now give a brief description of each session.

 \subsection{\HTHCOP  and \DCOP \ map from January 1997}
  In January 1997, we observed  simultaneously
  \HTHCOP (1--0), \DCOP (2--1),
 and \DCOP (3--2).  The corresponding half power beamwidths were 
28\arcsec , 17\arcsec , and 11\arcsec \ at 3, 2, and 1.3mm, respectively.
We observed a map of 67 positions with a grid spacing of 20\arcsec \
 relative to our reference position
 R.A.(1950) = $05^{h} 01^{m} 11.0^{s}$, Dec.(1950) = $25^{\circ}
 07^{'} 00^{"}$. We observed in frequency switching mode with a
 frequency throw of 1.0, 1.5, and 2.0 MHz at 3, 2, and 1mm, respectively.
The pointing was checked at 1-2 hourly
 intervals by means of a 3mm continuum scan
 on the nearby quasar 0528+134 and is thought to be good
 to $\sim$ 4 \arcsec . The relative pointing  of the three receivers was
 checked observing Mars  and was found to be accurate to within 2\arcsec.

 The backend for these observations was the facility autocorrelator
 split into three parts of bandwidth 20, 20, and 40 MHz at 3, 2, and 1mm,
respectively. The corresponding
 spectral resolutions were 0.033 \kms \ for \HTHCOP (1--0), 0.041 \kms \ for
 \DCOP (2--1), and 0.054 \kms \ for \DCOP (3--2).

\subsection{August 1997 Observations of \HCEIOP (1--0), \DTHCOP (2--1),
 \CSEO (2--1), \CEIO (1--0), \CEIO (2--1), and \CYAC (15--14)}
  In August 1997, we observed using two different set-ups. In
  both, we observed simultaneously at 3, 2, and 1.3 mm. In the
  first, we tuned to \HCEIOP (1--0), \DTHCOP (2--1),
  and \CSEO (2--1). In the second, we tuned to \CEIO (1--0), \CYAC (15--14),
and \CEIO (2--1).
  The basic observational procedure was as described above
  for the January observations. These observations  furnished a tentative
detection of \DTHCOP (2--1)  which was then re-observed in the
successive observing run.  The \CSEO (2--1) data were of poor quality due
to the high humidity and were discarded.
The \CYAC (15--14) line was detected and was further mapped in
our July(2000) measurements (see below).  \CYAC \ data will be presented
in a separated paper (\cite{CCW01}).

  \subsection{November 1998 observations of \HNFINP (1--0), \HCSEOP (1--0),
  \DTHCOP (2--1),  \CSEO (1--0), \NTDP (2--1), \CSEO (2--1)}
   In November 1998, we observed the 1--0 transition of
 (i)  \HNFINP \ and (ii) \HCSEOP (1--0)
 together with the 2--1  transition of \DTHCOP .
   In a second set-up, we observed the 1--0 and 2--1 lines of
   \CSEO \ together with the 2--1 transition of \NTDP .
The 1.3 mm data had to be discarded for technical reasons.
The 1--0 transition of \HNFINP \ was not detected  at offset (20\arcsec ,
-20\arcsec ) down to a limit of 0.047 K.

\subsection{July 2000 Observations of \NTHP (3--2), \NTDP (2--1), and
\NTDP (3--2)}
On July 8--10, we observed L~1544 in \NTHP (1--0), HC$_3$N(15--14),
\NTDP (2--1), \NTDP (3--2), and \NTHP (3--2).  The spectral
resolutions were as follows : 0.038 \kms , 0.051 \kms, and 0.042 \kms 
at the frequencies of \NTDP (2--1), \NTDP (3--2), and \NTHP (3--2), 
respectively. System temperatures (corrected for the
atmosphere) were $\sim$120 K for \NTHP (1--0), $\sim$250 K for \NTDP (2--1), 
$\sim$350 K for \NTDP (3--2), and $\sim$600 K for \NTHP (3--2) observations.  
The HPBW was 16\arcsec \ for \NTDP (2--1),
10\arcsec \ for \NTDP (3--2), and 9\arcsec \ for \NTHP (3--2).

\section{Observational Results}
\label{sobs_res}

 In this section we give an overview of our observational results.
We first consider the line profiles observed towards our (20,-20)
offset.  This is close to the high density peak in the
L~1544 discussed by WMA and by BAP.  We will refer to it in the
following as the dust emission peak. We then present our
maps of integrated intensity in \HTHCOP (1--0), \DCOP (2--1), \DCOP (3--2),
\NTHP (1--0),\NTDP (2--1),
and \NTDP (3--2) and compare with the dust emission maps.  These
comparisons allow us to extend and confirm the conclusion of \cite{CWT99}
(hereafter CWT99)
that whereas CO isotopomers are underabundant in the high density
peak of L~1544, certain ionic species remain relatively abundant.

\subsection{Spectra towards the dust emission peak}
\label{speak}
 In Fig.~\ref{fspectra}, we present spectra towards the (20,-20) offset
 in L~1544  and in Table~\ref{lpar} we
 present gaussian fit results for these spectra.  The form of these
 spectra has some importance for the later discussion and we
 therefore lay some emphasis on it here.  In the first place,
 as has been discussed by many authors (e.g. ~\cite{TMM98},
 \cite{WMW99}, CWT99), many of the observed profiles are
 double--peaked  and as a consequence are fit by two gaussian
 components in Table~\ref{lpar}.  This  could {\it a priori} be interpreted
 as implying two layers with different velocities along the line
 of sight but
 another interpretation is that, at least for the more
 abundant species, the observed profiles are affected by
  absorption in a low density ($10^4$ \percc )
  foreground layer which is
  at red--shifted velocities  (0.08 \kms )
  relative to the high density core
  (see e.g. ~\cite{WMW99}).

 If this latter view is correct, one expects  to see the effects of
 foreground absorption in low excitation lines of abundant
 species. Optically thin lines should show weaker single
 peaked features. However, our data show clearly that even
 weak lines of low abundance species
 such as \DTHCOP (2--1), \HCEIOP (1--0), \HCSEOP (1--0)
 (see ~\cite{DCC01}), and \DCOP (3--2)
 appear double with component separations between 0.24 and
 0.27 \kms \ and individual component widths between 0.2 and
 0.27 \kms. It is also important that
 there are slight offsets of order 0.05 \kms \ (15 kHz at 3mm)
 between the measured velocities for different species.
 It seems possible to us that this is due to the uncertainties in
 the spectroscopic constants but this requires verification.
In the case of \HCSEOP (1--0), the uncertainty in the measured
frequency is 20 kHz (Dore et al. 2001), corresponding to 0.07 \kms , 
which accounts for the observed velocity shift between this line
and \HCEIOP (1--0) (see Fig.~\ref{fspectra} and Tab.~\ref{lpar}).
 Thus  either there are two high density 
 velocity components along the line of sight  (and the
 transitions mentioned above are optically thin) or there are
 very high abundances of \HCOP \ and \DCOP \ in the postulated
 foreground layer allowing foreground absorption even in
 \DTHCOP (2--1) and \HCSEOP (1--0). The former is in agreement with the 
conclusions derived from C$^{34}$S by \cite{TMM98} and seems to us more likely.

 Another reason is the recent detection of 
 the three hyperfine components of \HCSEOP (1--0)
 at 87057.23 MHz (\cite{DCC01}).
  It is noticeable that the
 form of the \HCSEOP (1--0) profile (albeit with moderate S/N)
 is similar to that of \HCEIOP (1--0) as  one would expect if both
 were optically thin. Indeed, the intensity ratio is equal
 to 3$\pm$1 and 5$\pm$1 for the blue and red components, respectively,
 close to the [$^{18}$O]/[$^{17}$O] abundance ratio in the
 local interstellar medium (= 3.65, \cite{P81}).
 We conclude from this that \HCEIOP (1--0) is likely to be optically
 thin and thus that the double nature of the profile is in this
 case due to the existence of two velocity components along the
 line of sight.  Thus, foreground absorption seems
 improbable in this case.

 This does not imply that ``self--absorption effects'' are
 unimportant. On the contrary, our profile of \HTHCOP (1--0) is
 clearly broader with a larger separation between peaks
 (0.36 \kms ) than in \HCEIOP (1--0) and it is probable
 that \HTHCOP (1--0) is optically thick. And there are
 differences in the profiles of different hyperfine
 components of \NTHP (1--0)  which suggest to us that while
 most components are self--absorbed, the weakest
 $F_{1},F\, = \, 1,0\rightarrow 1,1$ component
 is close to being thin ($\tau_{1,0-1,1}$ $\la$ 0.7).
 For this reason, we sometimes have used the
 observations of this component  to derive
 column densities of \NTHP \, (see Paper II).
In general also, we conclude
 that observations of  weak   $< 1$ K lines of low abundance species are
 optically thin and can be used for column density determination.
 They appear often to be double--peaked towards the high density
 core of L~1544 and this is most simply interpreted as implying
 two separate velocity components along the line of sight.  In 
Section~\ref{kmodel} we discuss these components as arising from a 
combination of contraction and depletion.

 \subsection{\NTHP \ and \NTDP \ spectra}
  We discuss separately our July 2000 spectra of \NTHP (3--2) and
  \NTDP (2--1) and (3--2)  because there are few observations
  of these transitions available to date in such cold starless cores.
  Moreover, we expect these high J transitions to sample preferentially
  regions of high density within the L~1544 core.
  Interpretation of the spectra is complicated by the hyperfine stucture
  of these transitions. \NTHP (3--2)  for example has 45
  components spread out over roughly 6 \kms . \NTDP (2--1)
  has 38 components spread over 10 \kms . Although many of
  these components have negligible intensity, most of the structure which we
  observe in Fig. ~\ref{n2hpdpspectra} is due to hyperfine splitting.
  This is an advantage in that it permits a direct determination of
  optical depth based on the satellite line intensities but
  it makes it extremely difficult to infer the line profile in
  the region of line formation.

  We demonstrate this by  showing in Fig. ~\ref{n2hpdpspectra} our
  observed profiles at the molecular peak and the fits which we have
  made to the hyperfine structure.  The frequencies used for the fit
  are discussed in appendix~\ref{afreq} (see also the recent discussion
  of ~\cite{GPRFP01} for \NTDP ). These fits assume constant excitation
  temperature for all of the hyperfine--split transitions and estimate the
  total optical depth of a hypothetical unsplit transition.
  Table ~\ref{n2hpn2dp_par} summarizes the line parameters at offset
  (20,-20).  We note that \NTHP (1--0) and \NTDP (2--1) are
  clearly optically thick and we exploit this fact in the column
  density determinations discussed in Paper II.

  Because of the fact that we expect these transitions to probe the
  dense portion of the L~1544 core, it is of interest to study their
  kinematics.  Motions such as infall might give rise to wings on
  the observed profiles (\cite{RHM92}) and rotation
  of the dense core might cause a velocity gradient in the central
  region.  The presence
   of blended  hyperfine satellites however renders discussion
  of the intrinsic line profile extremely difficult.

 \subsection{Integrated Intensity Maps}

  Another useful input to the interpretation of the structure of the
  L~1544 high density core is given by the integrated intensity maps.
  CWT99 have already presented evidence that CO is depleted at
  densities above $10^5$ \percc \ in L~1544.
 % It is noticable also that
 % the CO isotopomers do not show the double--peaked profiles seen in
 % other species towards the dust emission peak (see Fig.~\ref{fspectra}).
 CWT99 also showed  that the integrated intensity maps
  of at least some ionic species peaked near the WMA dust emission
  maximum.  This is confirmed and extended by the maps shown in
  Fig.~\ref{fintegrated} where one sees that \DCOP (2--1), \DCOP (3--2),
 \HTHCOP (1--0), and \NTDP (2--1) all peak close to the WMA 1.3mm maximum.
   This is most clear in \NTDP (2--1) for which the general features of
   our map are consistent with those of  the dust emission.  \NTDP \
   is however more compact and we find
   for example half-power angular dimensions of 52\arcsec $\times $
   30\arcsec \ in \NTDP (2--1) as
   compared to 80\arcsec $\times$ 40\arcsec \
   in the 1.3mm continuum map of WMA.
   The morphological similarities
   allow us to conclude that although CO and many other species are
   highly depleted in the vicinity of the dust emission maximum in L~1544,
   \NTDP \ and to a lesser extent \NTHP are  not.  Thus one can expect that
   \NTDP \ traces  higher density material than CO and other
   depleted species. Indeed, \NTDP \ should probe
   the material within the region where the density
   ( deduced from dust emission) becomes roughly constant.

 On the other hand,  we note that
  \HTHCOP (1--0) is much more extended
  than \NTDP \ with a half--power size of  order
  100\arcsec . Thus if we consider \HTHCOP \ to be a tracer
  for the presence of CO in high density gas and \NTHP \
  to be a tracer for \MOLN , we conclude
  that CO is  more depleted than \MOLN \ at high densities.
 In both cases, the ions have higher dipole moments than
 the parent species (CO or \MOLN ) to which they are linked
 chemically. They can thus trace the presence of these species
 in situations where direct detection is impossible (in the case
 of CO because of the effect of foreground and background emission).
   The fact that \HTHCOP \ is much more extended than either the
   dust emission or \NTDP \  is a sign that CO is even more depleted
   in the central regions of L~1544 than would be implied by the
   observed \CSEO \ column density.

  We point out also that while the dust emission appears
  best traced by  \NTDP , some other species have a rough
  correlation with the dust peak.
  \DCOP \  peaks at  a position slightly offset to the NW but
  has similar half--power contours to the dust.
  Several other species (\CYAC \ and \AMM ,
  \cite{CCW01}, \cite{TMC01}) behave in analogous
  fashion.   CS on the
  other hand (like the CO isotopomers) does not (see ~\cite{TMM98}).
 There thus appear
  to be two "families" of species (dust-peak-phobic
  and dust-peak philic) which are depleted and (relatively)
  undepleted respectively in the high density core of L~1544.
  In general, ions and deuterated species are
  ``dust-peak-philic'' and thus survive to some extent in
  the high density material (see also \cite{BCL01}).  To what extent this is
  borne out by quantitative analysis
  is seen in Paper II.

  \subsection{Velocity field}
\label{svelocity}

We analysed the velocity structure in the observed maps following
the procedures in ~\cite{GBF93} to find a total velocity gradient
for each tracer.  We have also
 determined ``local gradients'' based upon 9-point maps
 for \HTHCOP (1--0), \DCOP (2--1), \DCOP (3--2), \NTHP (1--0), \NTDP (2--1),
 and \NTDP (3--2).  We request that at least 7 neighbouring points 
are available for the local gradient fit.  
These results are presented in Fig. ~\ref{vgrad}
 where the velocity  gradient
 vectors are superposed on the maps of integrated intensity
 shown in Fig. ~\ref{fintegrated}. 
 
 One sees in the first place
 that the observed velocity gradients indicate 
 quite a complex velocity field in the high density core
 of L~1544. Also, there are 
  distinct differences between the
 kinematics of different species with \NTDP (2--1) and \NTDP (3--2)
 having gradients predominantly along the minor axis  of L~1544
 while for other species, the situation is more complex. 
 In the case of \NTHP (1--0), one sees that towards the west,
 vectors are mainly along the minor axis parallel to those
 observed in \NTDP (3--2) but towards the SE and E, the vectors
 are along the {\it major} axis. We believe that these same
 structures are being seen superposed in the \HTHCOP \ and \DCOP \
 maps. However, there appears to be yet another feature on the 
 NE of the \DCOP (2--1) map with vectors along the minor axis but
 in the opposite direction to that seen in \NTDP (3--2). It
 is also noteworthy that the overall gradient seen in \DCOP (2--1)
 of 1.2 \kms ~ pc$^{-1}$ along the major axis
 is rather similar to that reported by 
 ~\cite{OLW99} using CCS. This suggests
that \DCOP \ and CCS trace the same regions within the core.
 However, it seems doubtful to us that overall rotation is being
 traced as the gradient does not seem to be continuous.  We
 conclude that 1.2 \kms pc$^{-1}$ should be treated as an upper
 limit to the overall rotation rate.  It is interesting to note that the 
direction of the total gradient as observed in \DCOP (2--1) and  \NTHP (1--0)
lines is similar to the direction of the magnetic field as measured by 
\cite{WKC00}.

 It is a striking feature of our results that the lines which
 best trace the dust peak region (see Fig. ~\ref{fintegrated}) have
 their principal gradient along the minor axis of L~1544. If one
 assumes that the material close to the dust emission peak is that
 with the highest density, this implies that the high density
 gas has large velocity gradients along the L~1544 minor axis. Moreover,
 the velocity gradients seen in \NTDP (3--2)  are the largest         
 which we have detected suggesting that the high density gas is
 moving faster. It should be realised however that there are some
 {\it caveats} which must be applied to the above discussion. One of
 these is the errors in observed velocities due to radiation transport
 effects such as the foreground absorption discussed earlier.
 %that as discussed in section ??, some of our observed profiles
 %are double peaked and many are clearly non--gaussian. This can be
 %due to unresolved relative motions in the high density material
 %or due to foreground self--absorption.  Nevertheless, we conclude
 This however is unlikely to affect the high 
 density material traced particularly well by 
 \NTDP (3--2)  where
 velocity gradients are as large as 7 \kms pc$^{-1}$ and
 along the minor axis.  An interesting feature of the \NTDP (3--2) 
emission map is its extension in the direction of the corresponding velocity
gradient, in the opposite sense to that of the other line maps, in 
particular the 2--1 of the same species.

 In a model where L~1544 has a ``disk--like'' structure seen almost
 edge--on, it is natural to attribute
 gradients along the minor axis  to infall
 and gradients along the major axis to rotation. Indeed,
 ~\cite{OLW99} claim  evidence for both based on  their CCS maps.
 The interesting point about our data however is that \NTDP ,
 whose integrated intensity correlates well with the dust emission
 shows {\it only} evidence for infall.  This suggests angular
 momentum loss between the  scales where \DCOP \ is
 abundant and those traced by \NTDP .  We note also that the
velocity gradients we estimate along the minor axis are
considerably larger (between 3 and 7 \kms pc$^{-1}$ in \NTDP \,
than in CCS and \DCOP along the major axis).  If these latter
estimates reflect real radial infall in the plane of the disk
(inclination 16$^{\circ}$ according to CB00), they are affected
by foreshortening and thus the gradient in the plane of the
disk is between 0.9 and 2.1 \kms pc$^{-1}$.  

An alternative interpretation of the velocity pattern  
seen in \NTDP (3--2) might be rotation of a prolate core
about its major axis.  Although this is a possibility, it seems 
to us unlikely because this gradient, in the minor axis direction, 
is only seen on the small scales 
($\sim$ 30\arcsec)  probed by \NTDP . At larger scales, 
the data indicate (if anything) a gradient along the 
major axis.  Conservation of angular momentum might be expected
to produce similarly oriented gradients in different density regimes.
However, we cannot presently exclude the possibility of rotation
in the central high density region, 
given the complex kinematics shown by the local 
gradients in Fig.~\ref{vgrad}.

\subsection{Variation of line widths across the core}
\label{swidth}
 Another tracer of  the velocity field in cores such as
 L~1544 is the observed line width. The
 discussion of the previous section had essentially to do
 with gradients in the centroid velocity in the plane
 of the sky. The measured line width in principle contains
 information about gradients in the line of sight of the centroid
 velocity. However, such effects are usually
 confused by both thermal line broadening and broadening due to
 local micro--turbulence and it is often difficult to distinguish
 the different contributions to observed profiles. In the present
 case however, the evidence that many observed profiles are
 non--gaussian is a strong indication that there are gradients
 also along the line of sight. 

  In Fig. ~\ref{dvrad}, we show observed line widths of \NTHP (1--0),
  \NTDP (2--1), and \NTDP (3--2) plotted against the projected
  offset from the dust continuum peak.  These line widths 
 have been derived correcting for hyperfine blending
 and optical depth assuming emission from a layer of homogeneous
 excitation temperature and an intrinsically gaussian profile. 
 In the case of \NTDP , the hyperfine blending effects are sufficiently
 large that departures from a gaussian profile would be extremely
 difficult to detect and hence the line width is the only useful
 parameter of the profile which can be derived. In the case of
 \NTHP (1--0) however, one clearly observes individual satellites
 with double--peaked profiles as has been discussed by various
 authors (~\cite{WMW99}) but we have nevertheless fitted a
 single component for purposes of comparison with \NTDP .
 
 Fig. ~\ref{dvrad} shows that the observed widths in \NTHP (1--0)
 are rather similar to \NTDP (2--1) and \NTDP (3--2) consistent
 with the lines being formed in the same region. Both in \NTHP (1--0)
 and in \NTDP (2--1), we see evidence for a slight fall-off of line
 widths from values of order  0.3 \kms \ close to the dust emission peak.
 There is no evidence in these data for an increase at large distances
 from the core nucleus as expected in some models where 
 magneto--hydrodynamic waves are incident on the core exterior (see e.g. 
\cite{GBW98}). 
 On the contrary, there is evidence for increased line broadening towards
 the dust peak of L~1544.  This is in qualitative agreement with
 infall (contraction) of the nucleus of L~1544. 

Linewidths of \DCOP (2--1) and (3--2) are also plotted against the
projected offset in Fig.~\ref{dvrad}.  At large distances from the center,
\DCOP \ linewidths show more scatter than \NTHP \ and \NTDP \ linewidths,
 probably due to the more
extended morphology.  No correlations between \DCOP \ linewidths and the 
distance from the (20, -20) offset appear from the figure.  This may in part 
be due to the larger optical depth of \DCOP \ lines (which do not
present hyperfine splitting) compared with \NTHP \ and \NTDP , and in part 
to the differential depletion between the precursors CO and \MOLN \ (discussed
in Paper II) which prevents \DCOP \ to trace the dense and highly CO--depleted 
nucleus of L~1544 where infall motions are more prominent.  

\section{Models of the kinematics}
\label{kmodel}

        In this section, we present two models of the kinematics of L 1544,
based on  models of CB00 and of MZ.  
Each model assumes a flattened structure
of dense gas threaded by a perpendicular magnetic field, and that gravity
drives the gas motions. 
The CB00 model assumes that the flattened structure
is a cylindrically symmetric disk, and predicts the radially inward motion
of gas in the disk plane due to ambipolar diffusion.  The MZ model
considers the quasistatic vertical contraction of the layer due to
dissipation of its Alfv\'enic turbulence. We note that polarization 
measurements at 
850 $\mu$m across L1544 (\cite{WKC00}) are in apparent disagreement with the 
picture adopted here (i.e. the magnetic field in the plane of the sky
seems to be roughly parallel to the major axis), although they show some angle 
deviation and some depolarization toward the core center. 

       The MZ and CB00 models treat what might possibly be earlier and later
stages of the same process, but they differ in their physical formulation
and in their adopted parameters. MZ consider the pressure of Alfv\'en waves
(\cite{MZ95}) as a key element in the vertical support of the
layer and the dissipation of these waves as the basis of the condensation
motions. Their layer is magnetically critical, and they neglect the slower
motions due to ambipolar diffusion in the plane of the layer.  In contrast,
CB00 consider a disk with no turbulence and a weaker field:  the disk is
magnetically supercritical by a factor greater than 2 for the conditions
adopted here within the central beam radius (12\arcsec \ or 0.008 pc) 
at time 2.7
Myr.  The CB00 disk is centrally condensed in the horizontal plane while the
MZ layer is horizontally uniform. The angle between the vertical axis of
the model and the plane of the sky ($\theta$) is 16$^{\circ}$ in CB00, 
estimated from the 
observed aspect ratio of the 1.3mm continuum map, and is chosen here to be 
75$^{\circ}$ for good fit between our \HCEIOP (1--0) data towards the (20,-20) 
offset and the MZ model. 
Thus in the present comparison the CB00 disk has nearly "edge-on" horizontal 
motions while the MZ
layer has more nearly "pole-on" vertical motions.  In the MZ model,
the velocity gradient observed along the minor axis (Sect.~\ref{svelocity})
may be interpreted as rotation of the layer about an axis in the
plane of the sky.
We use CB00 
results to predict line profiles of \NTHP (1--0) across the core and compare
  with observations.  

        We first explain how we derive "model profiles" and then compare
models with observations.

  \subsection{CB00 Synthetic Profiles}
   For purposes of simplicity, we have made a number of assumptions.
 In the first place, we have chosen here to present results based
 upon  the particular model of CB00 which gives best 
 agreement with the observed intensity distribution of mm and submm
 continuum (see also ~\cite{ZWG01}). This is the model at time $t_{3}$
 = 2.66 Myr when the central density is n(\MOLH) = $4.37\, 10^6$ 
 \percc . The density distribution in the mid--plane of this 
 structure has been given by CB00 and it is supposed
 isothermal at a temperature of 12~K (though the discussion of
 ~\cite{ZWG01} and \cite{ERSM01} makes it clear that this is quite
 a crude approximation). Then the density distribution perpendicular
 to the disk mid--plane can be computed assuming hydrostatic equilibrium.

 As far as the velocity field is concerned, we have again based our
 simulation on the results of ~\cite{CB00} for the radial (cylindrical)
 velocity in the disk mid--plane. We then assume that this holds also
 (in cylindrical coordinates) above and below the plane. One should note
 that this velocity field goes to zero at both the origin and at large
 radii with a maximum infall velocity (at time $t_{3}$) of 0.12 \kms \
 at radius 0.025 parsec . We have accordingly made a fit
 to the dependence of the radial velocity upon radius predicted
  for the ions at time $t_{3}$ by CB00 (their Fig.2) and
  assumed we can use this to estimate the column density of \NTHP \
  in a certain velocity range integrated along an arbitrary line of
  sight across the disk. 
  We thus can compute model profiles  by determining the
  hydrogen column density in a given  velocity range for all lines
  of sight across the model disk. 

  We have also run some models with a slight variation on the above
  procedure intended to simulate the fact that many species are
  depleted out in the high density nucleus of L~1544 (to a lesser extent
  even \NTHP \ is likely to be depleted).
  We assume a central
  ``hole'' in the density distribution for radii less than a critical
  value $r_{h}$ (which may differ from one species to another). 
  This has the consequence that lines of sight crossing the central
  hole have no contribution from the velocity range corresponding to
  the central region of L~1544. As one sees in Paper II, this is an
  extremely crude approximation but it is useful  in order to
  understand how one might obtain ``double--peaked'' profiles towards
  the center of L~1544. 

 We have carried out this procedure for $r_{h}=0.01$ pc (i.e. for
 the particular case of \NTHP \ )
 and present the resultant model
 profiles in Fig. ~\ref{mprof}. 
 Here we have convolved with a beam corresponding
 to 24\arcsec \ at a distance of 140 parsec and we show results for
 a disk inclined by 16$^{\circ}$  to the line of sight (i.e. close to
 edge--on).   We note that we consider a thermal broadening 
 when constructing these model profiles, assuming a gas temperature of
 about 10 K (as deduced from \CSEO , \CEIO , and \AMM \, observations; see 
Paper II and \cite{TMC01}), although spectra without thermal
 broadening are also shown in Fig. ~\ref{mprof} for comparison. On the
 other hand we
 neglect microturbulences. We also are implicitly
 assuming optically thin conditions.

 We see in Fig. ~\ref{mprof} that along the major axis of this
 contracting disk, we see symmetrical profiles with width of order
 0.2 \kms . There is also an indication of ``double peaks'', which does
 not go away in the central position even when thermal broadening is taken
 into account.    
 This is, we stress,
 due to the fact that we are viewing
 the model disk roughly edge on (in fact if we increase substantially
 the inclination, the indication for double peaks disappears and
 the line widths are reduced).  Along the minor axis on the
 other hand, profiles become asymmetric with blue shifts
 of roughly 0.1 \kms \ to the north (in the representation
 of Fig. ~\ref{mprof}) and correspondingly red shifts to the
 south. Clearly, one has symmetry about the major axis in this
 model whereas along ``E-W lines'' the profiles should not change. 
 As discussed in the previous section, one expects in this type of
 model velocity gradients along the minor axis as in fact observed in
 \NTDP .

 One final point worth noting with regard to this model is that it
 ``naturally'' produces line widths of the order of 0.2 \kms \ even
 without recourse to any local line broadening (see histograms
 in Fig. ~\ref{mprof}). Thus a significant fraction
 of the observed line width is probably due to systematic motions.
The model also predicts larger line widths towards the dust peak,
because of the larger infall speeds, and indeed this is observed
(Sect.~\ref{swidth}).

 \subsubsection{Comparison of CB00 model kinematics with observations of
 L~1544}
 \label{scomp}
  We here make a  comparison of predictions from our simulations
  of the CB00 model with observations.  When doing this,
  we have {\it not} made any systematic effort
  to fine--tune the models to improve
  the agreement with observations.  We have only varied
  the inclination at which the CB00 disk is viewed (as shown later).

In Fig. ~\ref{majmin}, we compare observed and model profiles 
  (with a central hole) along
  minor axes of L~1544 using the ``weak'' (hence presumed
  optically thin) $F_{1},F\, = \, 1,0\rightarrow 1,1$ satellite of
  \NTHP \ as a tracer.   The central hole has a radius of 0.01 parsec.  This
  latter value corresponds to around 2000 AU which is roughly the
  radius at which species such as \MOLN \ become depleted (see
  Paper II). One notes that the model profile at the (20,-20) offset 
has a clearly marked central dip, and that observed profiles have the same 
general characteristics as those of  the model, in the sense that there is 
evidence for  asymmetries to the blue (40,0) and red (0,-40) (the 
first moment of the \NTHP (1--0) intensity distribution changes by +0.06
\kms \ going from (40,0) to (0, -40)). However,
the observed profiles
  are wider than those predicted by the model.  

Model and observations
  do not match along the NW direction of the major axis (not shown), 
where \NTHP (1--0) lines continue to show double peaks up to a projected 
distance of 57\arcsec \ (or 0.04 pc) from the center, probably
  due to the particular morphology of L~1544 which is likely to be more
  complex than the assumed cylindrical symmetric model cloud.  Indeed,
the unconvolved velocity profiles in Fig.~\ref{mprof} (see histograms) 
show two peaks along the major axis of the model cloud, although 
their separation is not sufficiently large to be detected after 
convolution with a thermal Gaussian profile (see solid curves). 
It is possible that a reduction of the assumed 
electron fraction in the CB00 model (by a factor of $\ga$ 3, 
in view of the results presented in 
Paper II) may improve the agreement with observations, but this 
needs verification. 

 Another way of looking at these data is to consider the mean
 velocity as determined from first moments of the intensity
 distribution. 
  In Fig.~\ref{vmap}, we compare the measured velocities as a function
  of offset along the major and minor axes of L~1544 with  
  predictions from the simulation.  Along the minor axis, we
  see that the models show a sharp gradient close to the
  center as one might expect intuitively for a radial infall
  model. This is most marked for intermediate inclinations 
  (37$^{\circ}$) since in pure edge--on models, one sees both
  red and blue--shifted gas with equal weight. 
  Far from the center of the core, the contribution of
  gas at large radius (and low velocity) becomes large and the
  expected line of sight velocity decreases.
  Along the major axis in contrast, we
  expect at all inclinations no discernible relative velocity. Any gradients
  that are present must be due to other effects such as rotation. 

  We see from Fig.~\ref{vmap} that while there is reasonable agreement
  between the expected and observed velocity variation along the minor
  axis for \NTDP , there are clearly discrepancies for other species. 
  And along the major axis to the SE, all species (including \NTDP )
  show red--shifts contrary to model predictions. One could
  imagine that this latter effect was due to rotation except that
  transitions such as \NTHP (1--0) also show signs of red--shifts
  along the major axis to the NW. 

 Finally, there is a tendency for model lines to get broader toward 
the core center (factor of 1.5 from a projected distance of 40\arcsec \ 
to the center, see Fig.~\ref{mprof}), in agreement with what observed
(Fig.~\ref{dvrad}), although predicted line widths are significantly 
narrower than observed.  

 Our general conclusion from this set of comparisons is that contraction 
 in the plane of an inclined disk as in the CB00 model is likely to be
 one element of a successful kinematical model.  The simulation
 which we have made is certainly very crude and needs to be refined.
 A more realistic treatment of the abundance variation  and
 molecular excitation would be one such refinement. 
However, we suspect also that
 a successful model will have a somewhat larger infall velocity
 and will take account of rotation.

\subsection{MZ synthetic profiles and comparison with observed data}
\label{smz}

Since the MZ model is one dimensional, we did not attempt to
use it to reproduce the variation of profiles along the minor axis.
        To compute the MZ line profile of \HCEIOP (1--0) in Figure~\ref{fmz} 
we specify the vertical structure of the molecular density, the velocity
dispersion, the settling speed, and the excitation temperature. 
We briefly summarize these
relations here.  We approximate the vertical density structure of the MZ
layer as
\begin{eqnarray}
n & = & \frac{n_0}{1+(z/z_0)^2}   
\end{eqnarray}
where $n_0$ is the central density and $z_0$ a characteristic scale
height, given by
\begin{eqnarray}
z_0 & = & \frac{(1+R)  \sigma_T^2}{(2 \pi m G N)}         
\end{eqnarray}
with $m$ the mean molecular mass, $G$ the gravitational constant, $N$ the total
column density, $\sigma_T$ the thermal velocity dispersion, and $R$ the squared
ratio of the nonthermal and thermal velocity dispersions at the midplane.
The central density is related to the column density by
\begin{eqnarray}
n_0 & = & \frac{\pi G m}{2 (1+R)} \left( \frac{N}{\sigma_T}
 \right)^2 .
\label{eno}
\end{eqnarray}
The molecular abundance $n_{\rm mol}/n$ is assumed to be depleted to zero 
in the central half of the layer, i.e. for $|z| < z_0$, and to be constant 
for $z \geq z_0$.

        The nonthermal motions in the horizontal plane are due to
Alfv\'en waves.  The settling speed of the quasistatically condensing layer, 
at height z and at angle $\theta$, is
\begin{eqnarray}
v & = & -\frac{z R \Gamma sin(\theta)}{1 + (3/2) R}
\end{eqnarray}
where $\Gamma$ is the midplane dissipation rate of the turbulent motions.
This rate depends on the physical basis of the dissipation, for which MZ
adopt an MHD turbulent cascade, terminating in ion-neutral friction.
In general the dissipation is a function of time.  Here for
simplicity we take $\Gamma$ to be a constant, $\Gamma_f$, to be determined by the
best fit of the model to the spectral line.  With this assumption, the
turbulence parameter $R$ decreases with time $t$ from its initial value $R_i$
according to $R = R_i exp(-\Gamma_f t)$.

        The radiative transfer is computed assuming plane-parallel
transmission of optically thin line radiation, in 8 layers from $z/z_0$ =
-7/4 to 7/4 in increments of 1/2.  The excitation temperature of the
\HCEIOP \ transition is modelled simply according to a two-level excitation model.

             To obtain the fit shown in Fig.~\ref{fmz} 
we used the parameter values given in Table~\ref{tmz}.  For the parameters
adopted in Table~\ref{tmz}, the depletion zone has extent 1400 AU.

        In choosing $N$ we were guided, via eq.~\ref{eno}, by the 
requirements that 
the central temperature and density match those adopted in Paper II 
for the high-density core of L~1544, 10 K and $\sim 1 \times 10^6$ \percc \
respectively.  The
line width was fit by iteratively adjusting $R$, $\theta$, and $\Gamma$.  
The line
shape and height were set by $N$ and on the combination of depletion and
contraction.

We found that the model line profile is double-peaked only if the
model has both depletion and contraction.  The profile is single-peaked if
the model has depletion but no contraction, or if it has contraction but no
depletion. 

        Although the MZ model fits the observed central profile well, 
we caution that its application to the present problem is simplified for 
convenience
so that the uniform dissipation rate $\Gamma_f$ is a fit parameter and is not
derived from a physical dissipation process.  For the present parameter
choices 1/$\Gamma_f$ is comparable to the free-fall time from position $z_0$ to
the midplane of the model layer.  Thus $\Gamma_f$ corresponds to dissipation
which is about as fast as is physically allowable.  More detailed modelling
will be needed to make a more meaningful comparison.

\section{Conclusions}
\label{sdiscussion}

 This study has had as its main aim to study the high density
 core nucleus of L~1544.  One important result is that, while many
 species including CO are depleted at densities of $10^5$ or more,
 \NTHP \ and in particular \NTDP \ are less affected. \NTDP \ 
  best traces the region where the dust continuum radiation peaks and
 hence the kinematics of \NTDP \ offer a guide to the dynamical
 behavior of the high density core. We were able to detect
 the 3--2 transitions of both \NTHP \ and \NTDP \ and do not see
 evidence for ``collapse'' in the observed profiles.  However,
 this is in good part due to the difficulties caused by the blending
 of hyperfine satellites.

 We see in several apparently optically thin lines a profile
 showing ``double structure'' towards the nucleus of
 L~1544.  While we cannot exclude the presence of ``self--absorption'',
 our data suggest that at high densities (above
 a few times $10^4$ \percc ), one observes two components
 with different velocities along the line of sight. Although
 there is no compelling proof, it seems likely to us that we are
 observing a foreground and a background layer of moderate
 density ($3\, 10^4$ to $10^5$ \percc ) separated by the highest
 density gas from which all heavy species (with the possible
 exception of \MOLN ) have depleted out. This relative motion
 between foreground and background could plausibly be the infall
 at a velocity of order 0.1 \kms \ discussed by CB00.

 We have attempted to test this idea by comparing both 
 observed profiles and mean velocities with those predicted by
 CB00.  The observations show that those species
 which best trace the dust peak  have a velocity gradient
 along the minor axis of L~1544.
 This feature of the data is reproduced by our simulations even
 if the real velocity field is clearly more complex than in
 the model. Part of the difference could be due to superimposed
 rotation but we regard the present evidence for this as
 unconvincing. 

  The model profiles  also show  double peaked profiles in
  the central position of the disk in rough agreement with those
  observed.  Here however, the interpretation is difficult
 because there is good evidence that the profiles
 of low excitation 
 transitions  such as some of the \NTHP (1--0) satellites
 are affected by the foreground
 absorption mentioned above. We have attempted to minimise
 the importance of such effects using optically thin tracers
 in order to isolate the kinematics
 of the high density gas  and
 get good qualitative agreement even if the model profiles are
 somewhat narrower than those observed. 
  It is important to realise that models of this type
  in general predict ``double--peaked'' profiles 
  of optically thin high density tracers along the 
major axes, in reasonable
 agreement with observations (see histograms in Fig.~\ref{mprof}). 
In the CB00 model
 this is due to the disk--like structure 
 seen almost edge on and contracting radially in the plane of the disk.   
Because of thermal line broadening,  the double nature is maintained only 
at the center and if a central hole due to depletion is present.   

Although the agreement between observations and CB00 model is not 
perfect, the above results in our opinion give support to the
 idea proposed by many authors (~\cite{SAL87}, ~\cite{MC99} and
 references therein) that ambipolar
 diffusion is one of the basic processes of star
 formation. This is backed up by the results in Paper II which
 show that the ionization degree in the core of L~1544 is
 sufficiently low that the ambipolar diffusion time scale
 and the free--fall time scale are of the same order. 

Our data have also been compared with the MZ model.  
In Section~\ref{smz} we saw that this model reproduces the line profile 
of an optically thin tracer observed towards the dust peak to within 
its noise.  The success in reproducing the observed \HCEIOP (1--0) line 
width is due to the inclusion of turbulent motions.  As in the case of the 
CB00 model, the double peak structure is the result of depletion which 
removes gas at central positions and thus at central velocities, contributing 
to the dip in the line profile.  Since the MZ
model is horizontally uniform, it cannot be used in its present form to
compare with the variation of line profiles in the plane of the sky.
Nonetheless, the present result demonstrates that contraction along field
lines toward a depleted high--density zone may provide a viable explanation
for the observed profiles, suggesting that dissipation of Alfv\'enic turbulence
is an important process for the dynamical evolution of prestellar cores. 

The MZ model predicts that the 
nonthermal motions (and thus linewidths) get smaller as the gas gets denser.   
This is not what L1544 shows
at its center (see Fig.~\ref{dvrad}).  On the other hand, the CB00 model 
shows some line broadening towards the center (see Fig.~\ref{mprof}),
although linewidths are still significantly narrower than observed.
From this we conclude that a combination of the two models is probably needed 
to best reproduce our data.  In particular, we speculate that
the densest part of 
L1544 has broader lines because it is in transition from 1D condensation along
field lines (MZ model) to 3D condensation (1D along field lines as above, 
plus 2D across field lines via fast ambipolar diffusion as in CB00 model).
The key ingredient which allows the ambipolar diffusion to approach the 
free-fall rate is the
depletion which reduces the ambipolar diffusion time by reduction of the ion
fraction.  This point will be illustrated in Paper II. 

%, either by reduction
%of the ion fraction, or by reduction of the mean ion mass, or both.  This
%point will be illustrated in Paper II.

\acknowledgements
 The authors are grateful to the referee, Neal Evans, for useful comments
and suggestions.  PC and CMW
 wish to acknowledge travel support from ASI Grants 66-96,
 98-116, as well as from the MURST project ``Dust and Molecules
in Astrophysical Environments''.  

\clearpage

\appendix

\section{Frequencies for \NTHP (3--2), \NTDP (2--1),
and \NTDP (3--2)}
\label{afreq}

  Frequencies for the \NTHP \ and \NTDP \ transitions studied
  here
  have been computed on the basis of the molecular constants
  determined by ~\cite{CMT95}. We assume that the quadrupole
  coupling constants for the inner and outer nitrogen atoms are identical
  for \NTDP \ and \NTHP \ and we neglect effects due to the
  presence of deuterium. This gives for \NTDP \ essentially the
  same hyperfine pattern as ~\cite{GPRFP01} tabulate for
  \NTDP (2--1) and (3--2). There are then 45 separate components
  of the 3--2 lines for the two species and 38 for \NTDP (2--1).
  These have been used to fit the data shown in
  Fig.  ~\ref{n2hpdpspectra} where one sees that the
  approximation used is adequate. The blending is such however
  that we cannot improve the estimates for the hyperfine
  splitting.  We have however checked that the rest frequencies
  used are accurate to within 0.1 \kms \ if one assumes the
  same velocity for \NTHP (3--2), \NTDP (2--1), \NTDP (3--2)
  as for \NTHP (1--0). We have also collected
  for reference in
  Table  ~\ref{n2hp32freq} the frequencies of the
 16  \NTHP (3--2) transitions
  with more than 1 percent of the total line intensity. One sees
  from
  this that there is roughly 10 percent of the line intensity in
  the other 29 components.

\begin{center}
 \begin{deluxetable}{cccccc}
 \tablewidth{0pc}
 \footnotesize
 \tablecaption{Frequencies for \NTHP (3--2)
 }
 \tablehead{
  \colhead{F$_{1}^{'} {\rm F}^{'}$} & \colhead{F$_{1}$ F} & \colhead{Frequency} &
  \colhead{Error} & \colhead{Rel. Velocity $^{a}$} &
 \colhead{Rel. Intensity}  \nl
  &  & \colhead{MHz} & \colhead{MHz} &  \colhead{\kms}
  &   }
 \startdata
 \hline
   3 4 &  3 4 &   279509.8785 &      0.0062 &       2.0120 &       0.0166   \nl
   2 2 &  1 2 &   279511.1328 &      0.0064 &       0.6690 &       0.0154   \nl
   2 2 &  1 1 &   279511.3445 &      0.0061 &       0.4424 &       0.0479   \nl
   2 1 &  1 0 &   279511.3847 &      0.0061 &       0.3993 &       0.0222   \nl
   3 3 &  2 3 &   279511.4305 &      0.0063 &       0.3503 &       0.0140   \nl
   2 3 &  1 2 &   279511.5089 &      0.0062 &       0.2663 &       0.0945   \nl
   3 3 &  2 2 &   279511.6858 &      0.0061 &       0.0769 &       0.0885   \nl
   3 2 &  2 1 &   279511.7978 &      0.0061 &      -0.0430 &       0.0610   \nl
   4 3 &  3 2 &   279511.8097 &      0.0061 &      -0.0557 &       0.1005   \nl
   4 4 &  3 3 &   279511.8114 &      0.0061 &      -0.0576 &       0.1356   \nl
   3 4 &  2 3 &   279511.8083 &      0.0061 &      -0.0542 &       0.1249   \nl
   2 1 &  1 1 &   279511.8486 &      0.0063 &      -0.0974 &       0.0187   \nl
   4 5 &  3 4 &   279511.8621 &      0.0061 &      -0.1118 &       0.1746   \nl
   4 3 &  3 3 &   279512.3171 &      0.0065 &      -0.5990 &       0.0102   \nl
   2 2 &  2 2 &   279514.2219 &      0.0063 &      -2.6385 &       0.0114   \nl
   2 3 &  2 3 &   279514.3427 &      0.0063 &      -2.7678 &       0.0141   \nl
 \enddata
 \tablenotetext{a}{Velocities are relative to an assumed rest frequency
 of 279511.757 MHz}
% \tablenotetext{}{References: (1) \cite{G82}; (2)
% Dore et al. (A\&A, submitted); (3) ??? ; (4) \cite{CMT95}
% (component $F_1,F$ = 0,1 $\rightarrow$ 1,2);
% (5) \cite{PP85}; (6) \cite{FL81};
% (7) electronic Lovas catalogue; (8) \cite{L92}; (9)
% Dore (priv. comm.) (component $F_1, F$ = 3,4 $\rightarrow$ 2,3)}
 \label{n2hp32freq}
 \end{deluxetable}
 \end{center}

\clearpage

\begin{center}
\begin{deluxetable}{lcccccc}
\tablewidth{0pc}
\footnotesize
\tablecaption{Line parameters from gaussian fits to spectra taken at the
(20,-20) offset.}
\tablehead{
 \colhead{Transition} & \colhead{Frequency} & \colhead{Ref.} &
 \colhead{$T_{\rm mb}$} & \colhead{$\int {{\rm T}_{mb}dv}$} &
\colhead{V$_{lsr}$} & \colhead{$\Delta v$} \nl
 & \colhead{MHz} & & \colhead{K} & \colhead{K \kms} & \colhead{\kms}
 & \colhead{\kms} }
\startdata
\hline
\HCEIOP (1--0) & 85162.222 & 1 & 0.35\p0.05 & 0.07\p0.01 &
	7.04\p0.01 & 0.18\p0.03 \nl
               &           &   & 0.39\p0.05 & 0.09\p0.01 &
        7.28\p0.01 & 0.23\p0.03 \nl
\HTHCOP (1--0) & 86754.279 & 1 & 1.7\p0.2 & 0.33\p0.03 &
        7.03\p0.01 & 0.24\p0.02 \nl
               &           &   & 1.1\p0.2 & 0.24\p0.03 &
        7.39\p0.01 & 0.20\p0.03 \nl
\HCSEOP (1--0)\tablenotemark{a} &  87057.258 & 2 & 0.06\p0.01& 0.011\p0.003&
	7.15\p0.03& 0.19\p0.06 \nl
               &           &   & 0.05\p0.01& 0.010\p0.003& 7.38\p0.03&
 	0.18\p0.07 \nl
\HNFINP (1--0)\tablenotemark{b} & 90263.833 & 3 & $<$ 0.047 & \nodata &
        \nodata & \nodata \nl
\NTHP (1--0)\tablenotemark{c} & 93173.2650 & 4 & 1.96\p0.05 & 0.83\p0.01 &
	7.154\p0.003 & 0.397\p0.006 \nl
\CEIO (1--0) & 109782.1734 & 5 & 4.9\p0.2 & 1.53\p0.03 &
        7.180\p0.002 & 0.292\p0.005 \nl
\CSEO (1--0) & 112358.988 & 6 & 2.19\p0.04 & 0.61\p0.02 &
       7.189\p0.001 & 0.274\p0.005 \nl
\DTHCOP (2--1) & 141465.141 & 7 & 0.22\p0.03 & 0.05\p0.01 &
       7.08\p0.02 & 0.20\p0.04 \nl
               &            &   & 0.11\p0.03 & 0.02\p0.01 &
       7.35\p0.04 & 0.20\p0.08 \nl
\DCOP (2--1) & 144077.321 & 3 & 3.1\p0.1 & 0.88\p0.06 &
        7.098\p0.001 & 0.271\p0.008 \nl
             &            &   & 1.0\p0.1 & 0.49\p0.06 &
        7.37\p0.03 & 0.46\p0.05 \nl
\DCOP (3--2) & 216112.623 & 3 & 2.4\p0.3 & 0.7\p0.1 &
        7.14\p0.02 & 0.27\p0.03 \nl
             &            &   & 0.6\p0.3 & 0.17\p0.09 &
        7.40\p0.05 & 0.26\p0.05 \nl
\CEIO (2--1) & 219560.3568 & 5 & 4.4\p0.8 & 2.0\p0.2 &
        7.19\p0.02 & 0.43\p0.04 \nl
\enddata
\tablenotetext{a}{Values refer to the main hyperfine
component.}
\tablenotetext{b}{Upper limit based on assumed width of
0.3 \kms .}
\tablenotetext{c}{Values refer to the component $F_1,F$ = 0,1 $\rightarrow$
1,2.}
\tablenotetext{}{References: (1) \cite{G82}; (2)
Dore et al. (2001); (3)  \cite{L92}; (4) \cite{CMT95};
(5) \cite{PP85}; (6) \cite{FL81};
(7) electronic Lovas catalogue; (8) Dore (priv. comm.)}
\label{lpar}
\end{deluxetable}
\end{center}

\begin{center}
 \begin{deluxetable}{lrccccc}
 \tablewidth{0pc}
 \footnotesize
 \tablecaption{Line parameters from fits to the hyperfine structure
 of the \NTHP (1--0), \NTHP (3--2), \NTDP (2--1), and \NTDP (3--2)
  spectra taken at the (20,-20) offset\tablenotemark{a}.}
 \tablehead{
  \colhead{Transition} & \colhead{Frequency} &
  \colhead{$\int {{\rm T}_{mb}dv}$} & \colhead{$T_{\rm ex}$} &
 \colhead{$V_{\rm LSR}$} &
 \colhead{$\Delta v$} & \colhead{$\tau $} \nl
  & \colhead{MHz} & \colhead{K \kms}& \colhead{K} & \colhead{\kms} &
  \colhead{\kms} & \colhead{} }
 \startdata
 \hline
 \NTHP (1--0) & 93173.4035\tablenotemark{b} & 5.49\p0.03 & 5.0\p0.1 &
	7.157\p0.001 & 0.311\p0.002 & 13.3\p0.4 \nl
	      & & 5.15\p0.01& 4.90\p0.09& 7.162\p0.001&
		0.304\p0.001& 13.8\p0.2 \nl
 \NTDP (2--1) & 154217.094\tablenotemark{c} & 1.45\p0.05 & 4.8\p0.5 &
	6.957\p0.004 & 0.285\p0.009 & 5.1\p0.4 \nl
              & & 1.72\p0.02 & 4.6\p0.3 & 6.968\p0.002 & 0.287\p0.005 &
		4.0\p0.2 \nl
 \NTDP (3--2) & 231321.9063\tablenotemark{d} & 0.49\p0.04 & \nodata &
	7.04\p0.01 & 0.31\p0.03 & 0.1\p0.2 \nl
	      & & 0.47\p.0.01 & 4\p2 & 7.050\p0.007 & 0.32\p0.02 & 1.4\p0.6 \nl
 \NTHP (3--2)  & 279511.9375\tablenotemark{d} & 0.45\p0.08 & \nodata &
	7.14\p0.02 & 0.18\p0.03 & 0\p2 \nl
	      & & 0.67\p0.04 & 4\p1 & 7.117\p0.009 & 0.16\p0.02 & 6\p2 \nl
 \enddata
 \tablenotetext{a}{Parameters in first rows refer to the spectrum at (20,-20),
whereas those in second rows are from the spectrum averaged over a
9$\times$9 grid of positions, spaced by 10$^{\prime\prime}$.}
 \tablenotetext{b}{Central line frequency from \cite{CMT95}.}
 \tablenotetext{c}{Frequency of the main hyperfine component ($F_1,F$ =
	3,2 $\rightarrow$ 2,3), blended in the main group of hyperfines.}
 \tablenotetext{d}{Weighted mean of hyperfines.}
 \label{n2hpn2dp_par}
 \end{deluxetable}
 \end{center}

\begin{center}
 \begin{deluxetable}{cccccc}
 \tablewidth{0pc}
\tablecaption{Parameters Used for MZ Model Line Profile\tablenotemark{a}}
\tablehead{
\colhead{$R$} & \colhead{$\theta$} & \colhead{$N$} & \colhead{$\Gamma_f$} &
 \colhead{$T$} & \colhead{$v_0$} \nl
 & \colhead{($\deg$)} & \colhead{($10^{22}$ cm$^{-2}$)} & 
 \colhead{($10^{-12}$ s$^{-1}$)} & \colhead{(K)} & \colhead{(\kms )}}
\startdata
\hline
0.8 &  75 &   4.4 &  1.8 &  10 & 7.16 \nl
\enddata
\tablenotetext{a}{$R$ is the midplane turbulence parameter, $\theta$ the
angle between the vertical and the plane of the sky, 
$N$ the total column density,
$\Gamma_f$ the best-fit constant turbulent dissipation rate, $T$ the kinetic
temperature, and $v_0$ the LSR velocity of the observed cloud.}
\label{tmz}
\end{deluxetable}
\end{center}

\clearpage

% FIGURE CAPTIONS

\figcaption[spectra.ps]{Spectra of detected lines toward offset (20, -20),
the ``molecular peak''.  From top to bottom:  \CEIO (1--0),
component $F$ = 5/2 $\rightarrow$ 5/2 of \CSEO(1--0), \HTHCOP (1--0),
\DCOP (2--1), \DCOP(3--2), component $F_1,F$ = 0,1 $\rightarrow$ 1,2
({\it thin line}) and $F_1,F$ = 1,0 $\rightarrow$ 1,1 ({\it dashed line}) of
\NTHP (1--0), \HCEIOP (1--0), \DTHCOP (2--1), and component
$F$ = 5/2 $\rightarrow$ 5/2 of \HCSEOP (1--0).  Dotted line mark the
$V_{\rm LSR}$ determined from the hfs fit of \CSEO (1--0)
(= 7.189 \kms, see Tab.~\ref{lpar}).  
\label{fspectra}}

\figcaption[l1544_n2hpdpspe.ps]{Histogram spectra of \NTHP (3--2),
\NTDP (2--1), and \NTDP (3--2) towards the molecular peak of
L~1544. An average has been made of the spectra towards offset
(20,-20) and at 8 positions offset by 10-15 arc seconds
relative to (20,-20) (i.e (10,-30) (20,-30) (30,-30) (10,-20) ...).
We also show (solid curve) fits to the hyperfine structure of
these lines made using the parameters discussed in the appendix.
\label{n2hpdpspectra}}

\figcaption[integrated.ps]{Integrated intensity maps of \DCOP (2--1),
\DCOP (3--2), \HTHCOP (1--0), \NTHP (1--0), \NTDP (2--1), and \NTDP (3--2) 
overlapped to the
1.3mm continuum emission map from WMA
smoothed at a resolution of 22$^{\prime\prime}$
({\it grey scale}). Contour levels are
30, 50, 70, and 90\% of the peak [1.3 K \kms, for \DCOP (2--1) at
offset (20, -20); 0.8 K \kms for \DCOP (3--2) at (20, -20);
0.8 K \kms for \HTHCOP (1--0) at (0, -20); 5.5 K for \NTHP (1--0) at 
(20, -20); 2.1 K \kms for \NTDP (2--1)
at (20, -20); 0.6 K for \NTDP (3--2) at (30, -20)].  
The peak of the 1.3 mm map is 225 mJy/22\arcsec at offset
(26, -21).  Note that all integrated intensity maps, in particular
the \NTDP (3--2) one, have a peak nearby
the dust peak, unlike the \CSEO (1--0) map (see CWT99). \label{fintegrated}}

\figcaption[gradient.ps]{Velocity gradient 
vectors in the plane of the sky derived
from our 30-m maps of L~1544. We show from the top left going
clockwise maps of \DCOP (2--1), \DCOP (3--2), \NTHP (1--0),
\NTDP (3--2), \NTDP (2--1), \HTHCOP (1--0).
The integrated intensity maps of Fig. ~\ref{fintegrated} are in the
background.
 The thick arrows at the
bottom right of each map shows velocity gradients averaged over the
map and the numbers give the magnitudes of the gradients in
km/s/pc. The thin arrows over the map show the magnitude (their length is 
proportional to the corresponding magnitude, in units of the total gradient
magnitude) and direction of
the velocity gradients derived from 9 neighbouring point regions across the
maps where $T_{\rm mb}/\sigma_{\rm T_{mb}} > 3$. 
Points show the measured positions. 
With the exception of \NTDP (3--2), which best traces 
the contracting high density L1544 nucleus, all the other lines show 
complex velocity fields which cannot be simply ascribed to solid 
body rotation and infall. 
\label{vgrad}}

\figcaption[dv_radius.ps]{Measured line widths in \NTHP (1--0),
\NTDP (2--1), \NTDP (3--2), \DCOP (2--1), and \DCOP(3--2) 
plotted against the distance from the L~1544 ``molecular peak''. 
The \NTHP \ and \NTDP \ measurements shown have been corrected for 
hyperfine blending.  Note the slight fall off of the \NTHP \ and 
\NTDP \ linewidths with distance from (20, -20) offset indicative 
of infalling motions towards the core center.
\label{dvrad}}

\figcaption[mod_prof.ps]{ Model profiles of an optically thin
satellite of the \NTHP (1--0) line shown as a function of map
offset for a model disk having the density distribution described
by CB00 for time $t_{3}$, with an inner hole of radius
$r_{h}$ = 0.01 pc (2100 AU). The velocity scale (in \kms ) is reported for the
spectrum at offset (40, -20). The solid profiles take into account the
line broadening due to thermal velocity dispersion for a temperature of about
10 K, while the histograms are purely kinematic.
The disk like structure is supposed inclined
by an angle of  16$^{\circ}$ to the line of sight and model profiles
have been convolved with a beam of 24\arcsec .  The x-axis of this 
figure is supposed to represent the major axis of the disk and
the y-axis the minor axis. Adjacent model spectra are separated by
20\arcsec .  The (0,0) position in this model map corresponds to the 
(20,-20) offset in Fig.~\ref{fintegrated}.  
\label{mprof}}

\figcaption[majmin.ps]{CB00 model (curve) and observed profiles
(histogram) of the \NTHP \,
$F_{1},F\, = \, 1,0\rightarrow 1,1$ satellite
along the minor axis of L~1544.
Model profiles are obtained with an inner hole as in
Fig. ~\ref{mprof}.
Offsets are given in arc seconds in the top right corner of
each panel. The normalization of the intensity scale is
arbitrary.  The peak of the observed hyperfine shows a shift 
from blue to red going from north--east to south--west across
the minor axis, correctly reproduced by the CB00 model, although 
model linewidths are narrower than observed.  
\label{majmin}}

\figcaption[velmap.ps]{Cuts  showing the
 mean line of sight  velocity derived from the model
profiles similar to those of Fig. ~\ref{mprof}  as a function of offset
on the sky.   Both for the models and data, we have used the
velocity determined from first moments of the intensity distribution.
The panels show the expected
and observed velocities for cuts along the major axis (right) and
the minor axis (left).  Measurements are shown by the error bars.
 Model curves for the minor axis 
 are plotted for inclinations of 70$^{\circ}$ (dotted),
37$^{\circ}$ (dashed) and 16$^{\circ}$ (full). For the major axis,
the CB00 model predicts no velocity offset (full horizontal line).
Note the good agreement between the model and observed velocity
profile along the minor axis for the \NTDP \ lines, which best trace the 
dense and highly CO--depleted nucleus of the core.  
 \label{vmap}}

\figcaption[phil.ps]{Comparison of MZ model line profile (smooth curve) and 
\HCEIOP (1--0) spectrum (histogram) from L1544 position (20, -20).  The MZ 
model is based on a self-gravitating layer which condenses quasistatically 
along magnetic
field lines as it dissipates its Alfvenic turbulence.  The model parameters
are described in Table~\ref{tmz} and are discussed in the text. The MZ
model reproduces the linewidth and the line profile substantially better 
than the CB00 model, although it cannot be used for studies of line 
variation across the core.
\label{fmz}}

\plotone{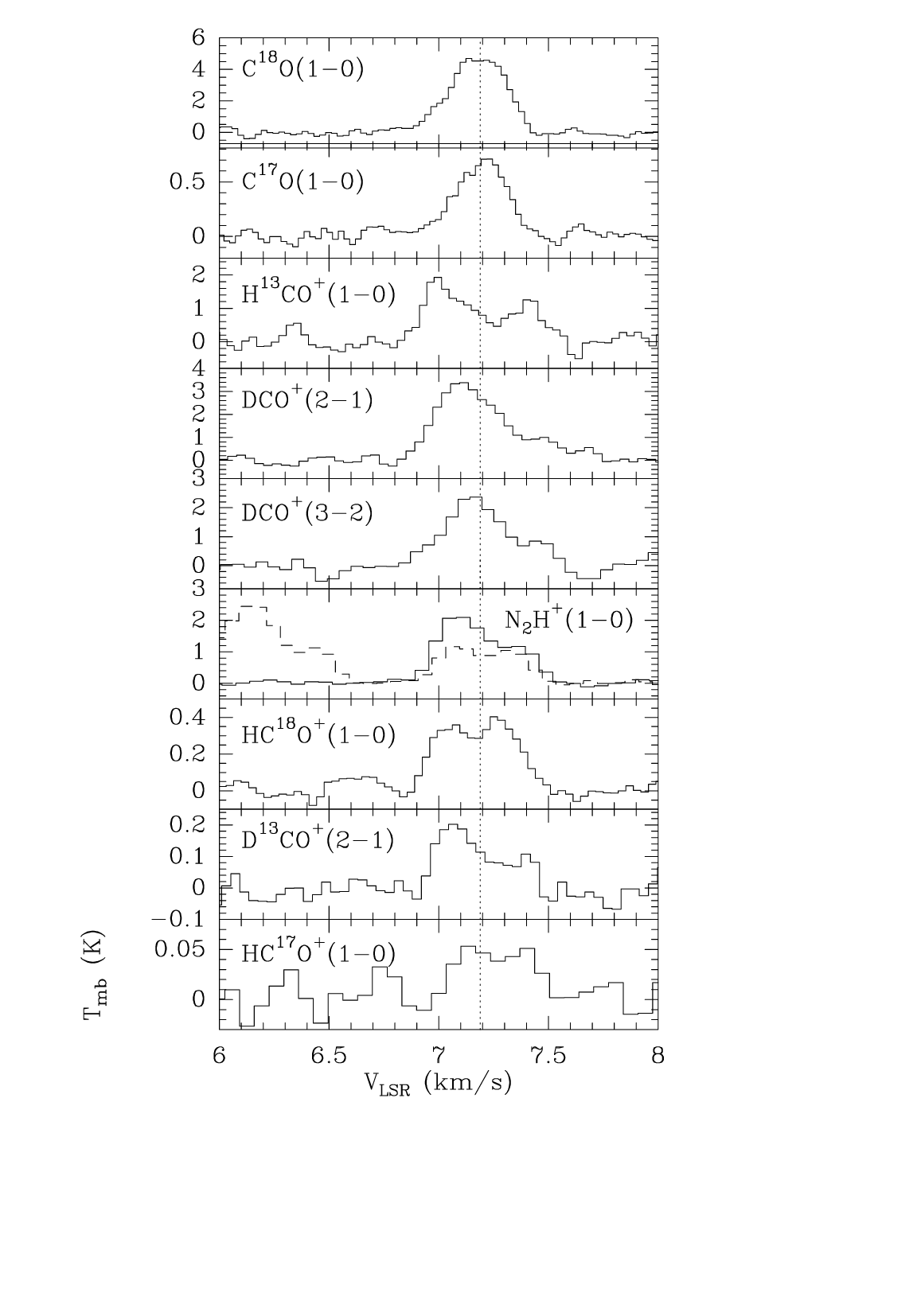}
Fig.1

\plotone{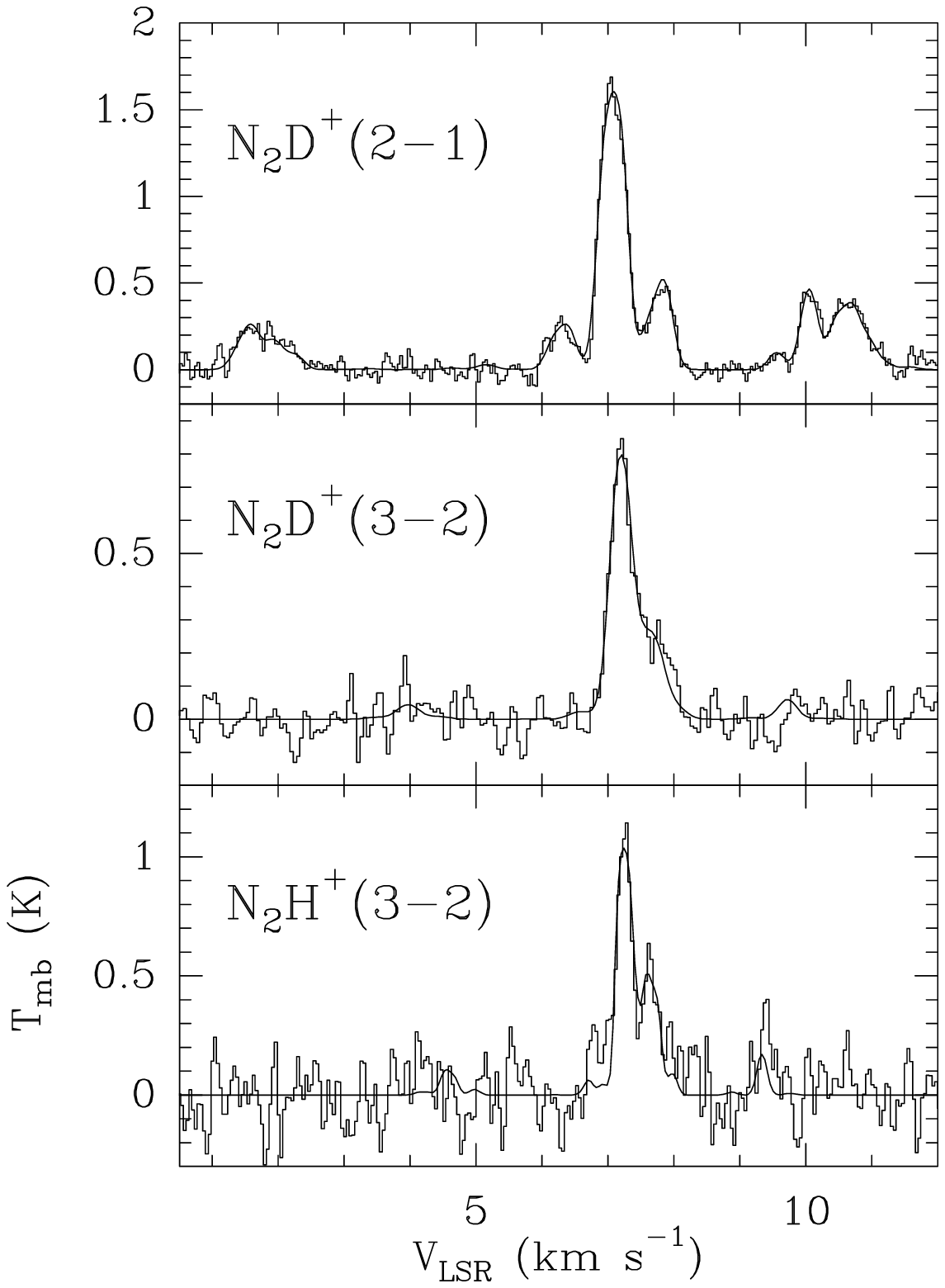}
Fig.2

\plotone{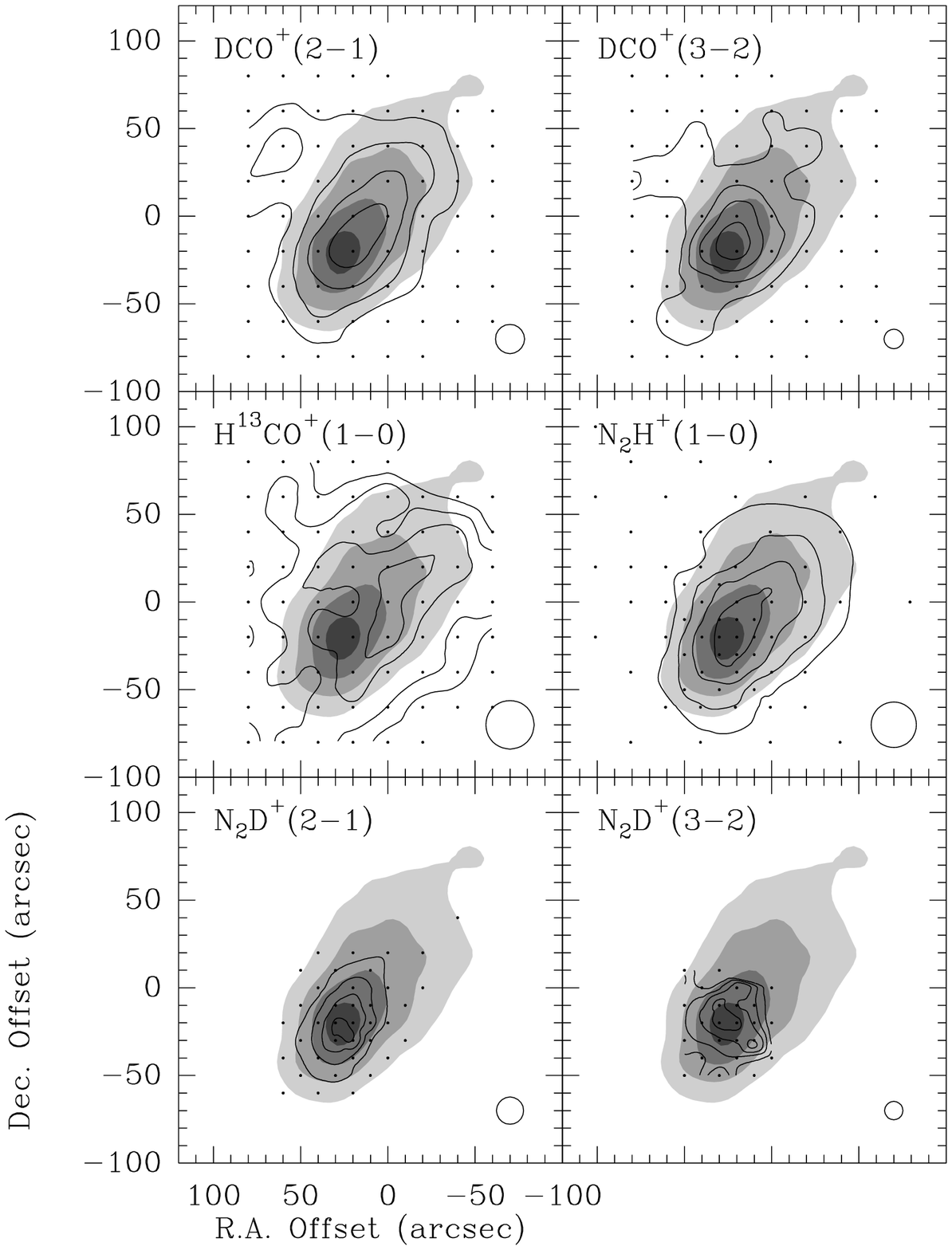}
Fig.3

\plotone{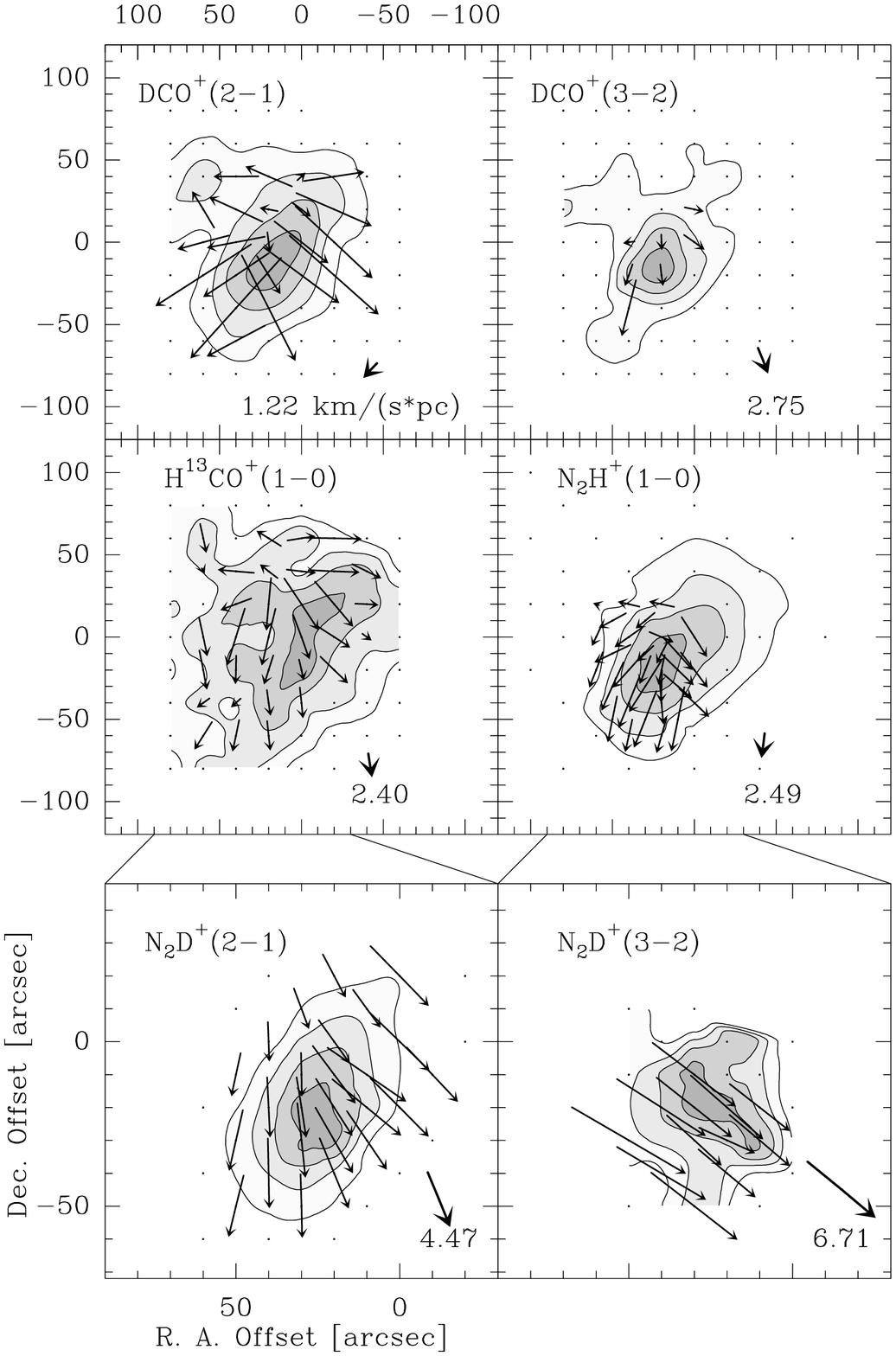}
Fig.4

\plotone{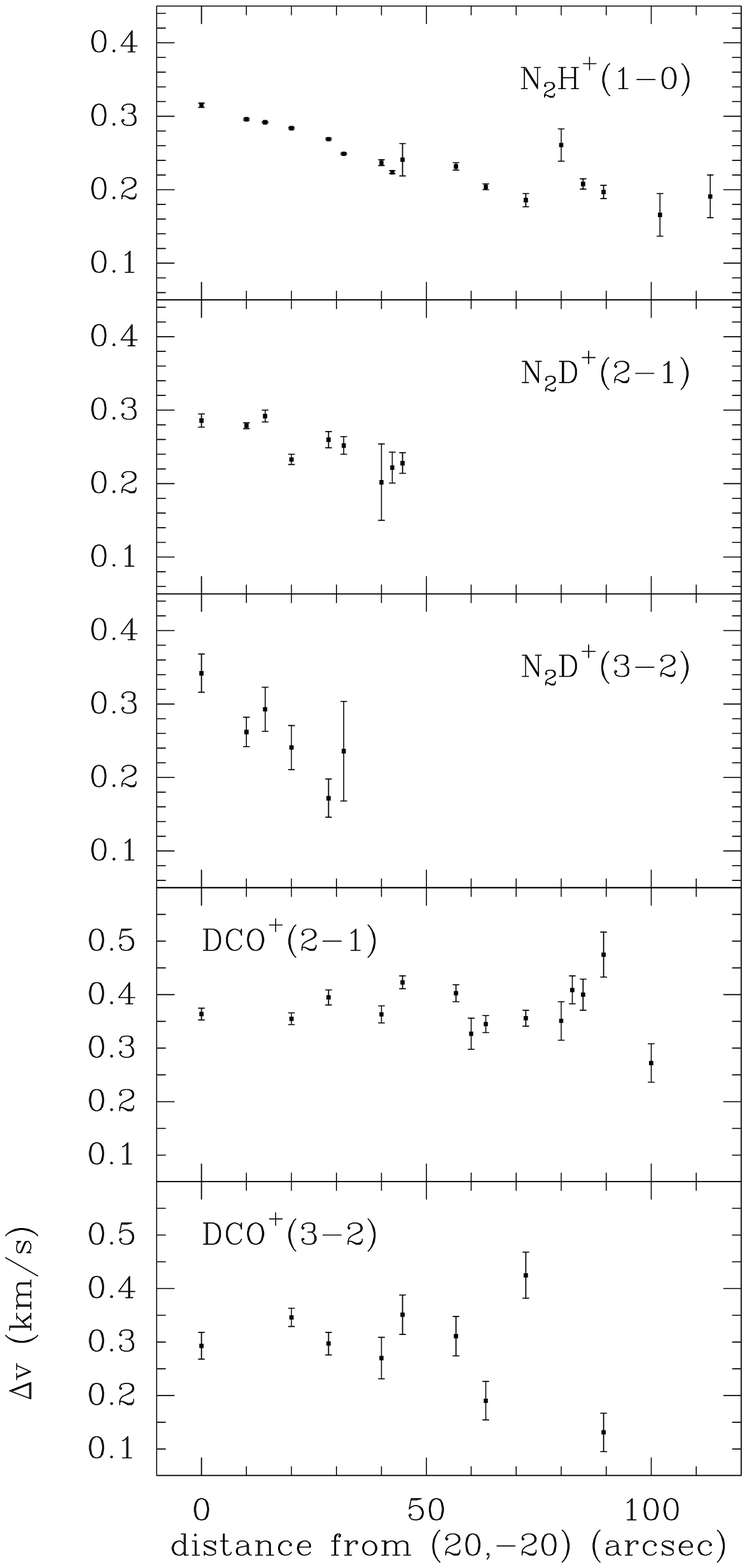}
Fig.5

\plotone{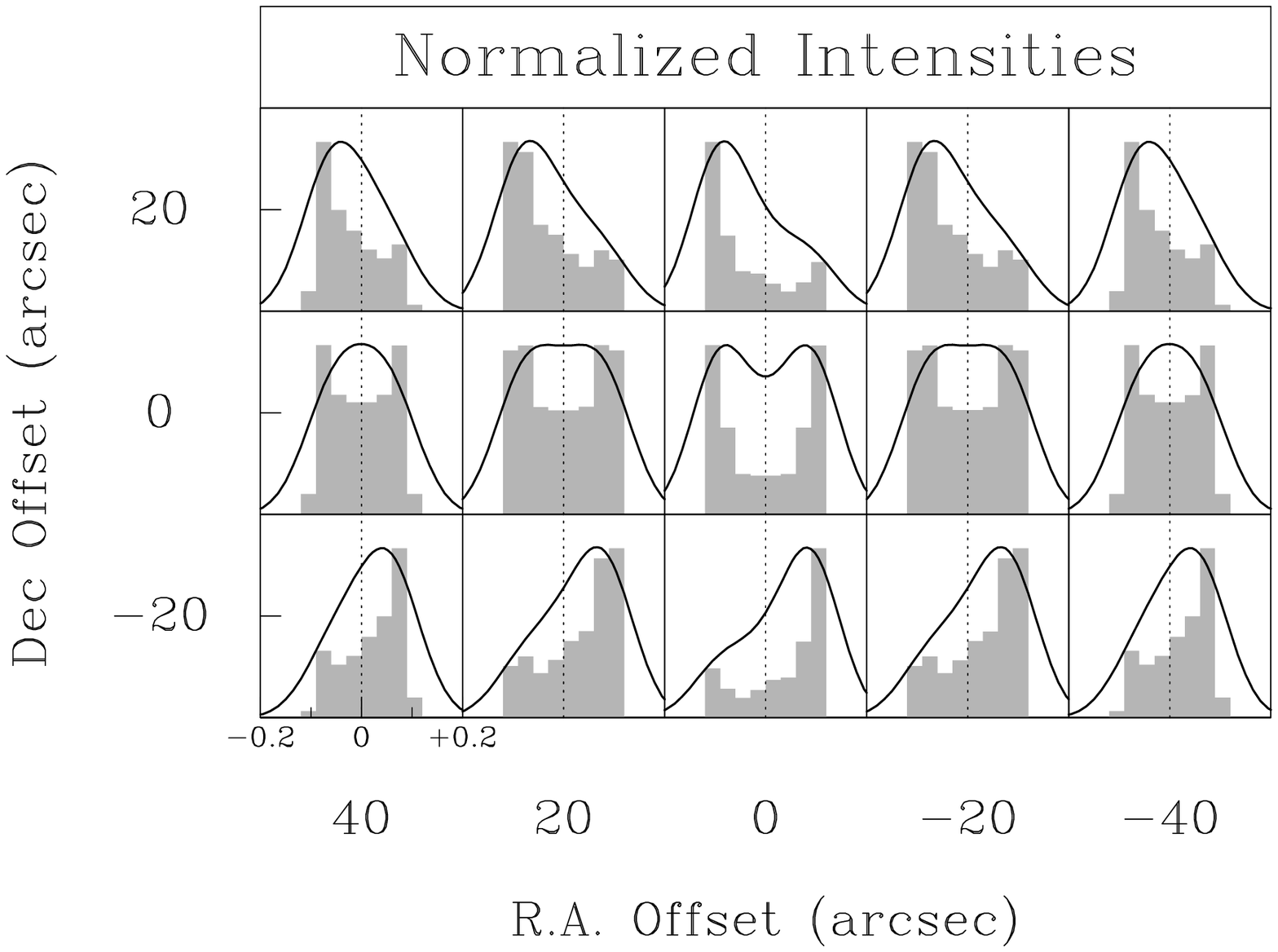}
Fig.6

\plotone{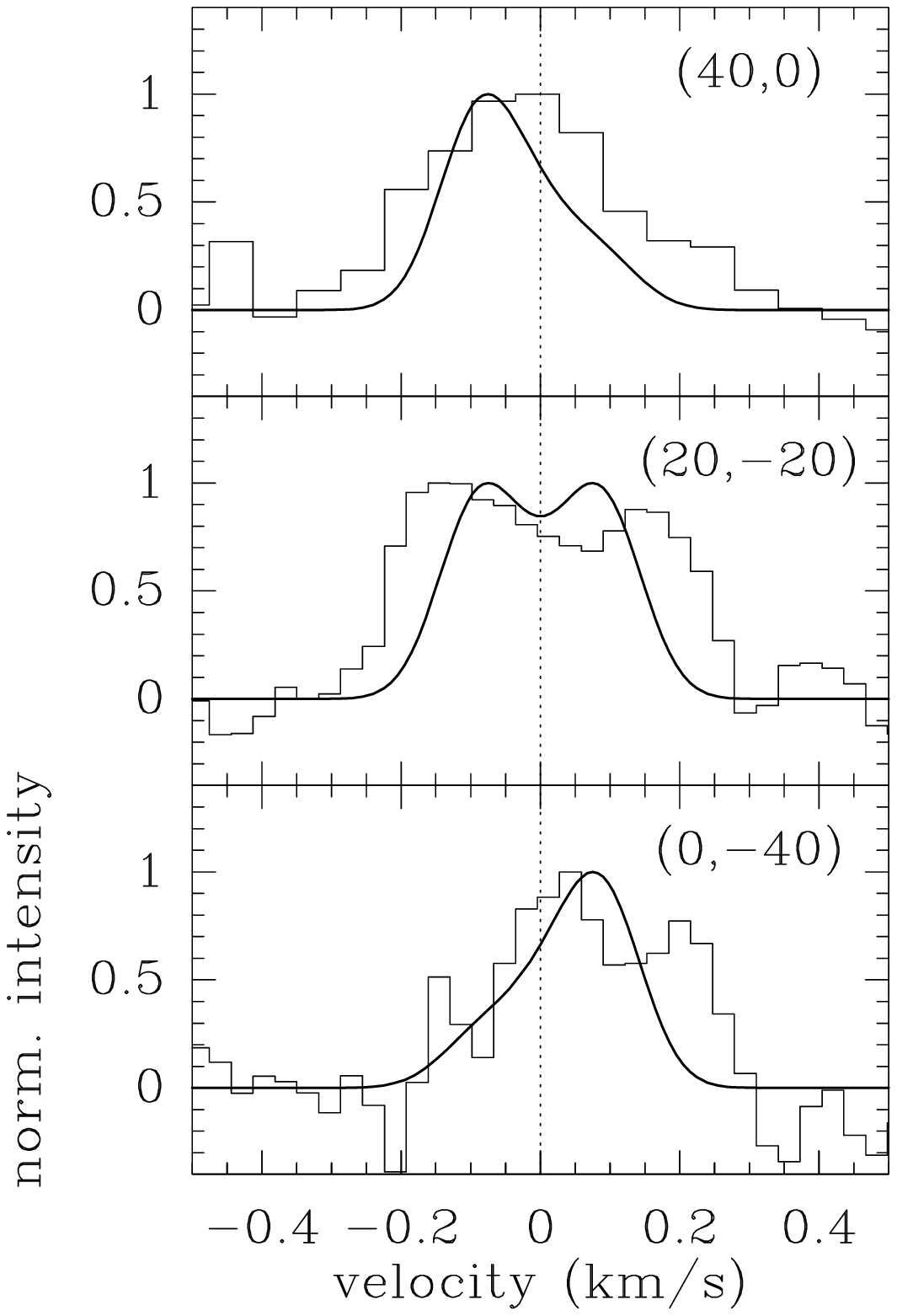}
Fig.7

\plotone{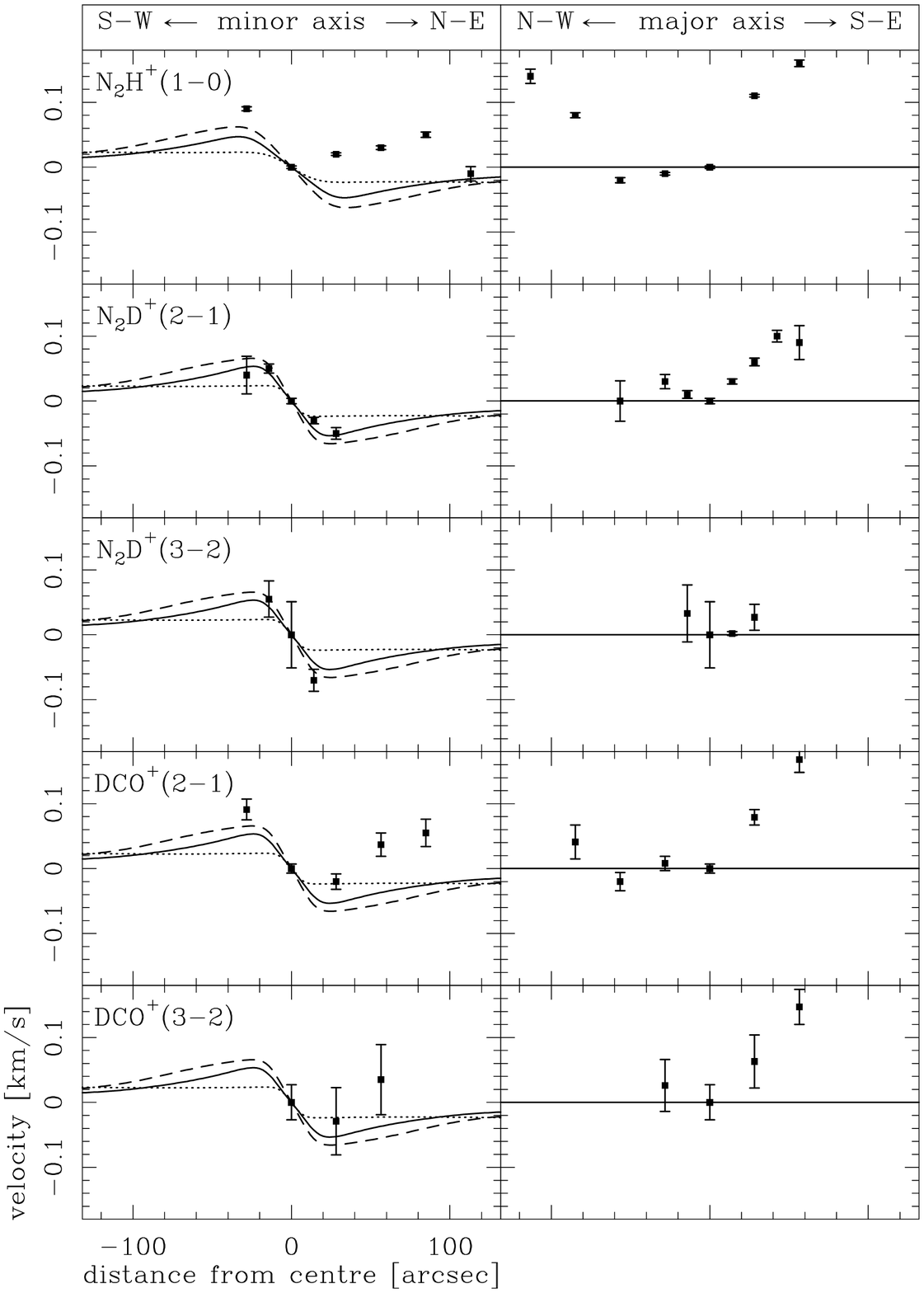}
Fig.8

\plotone{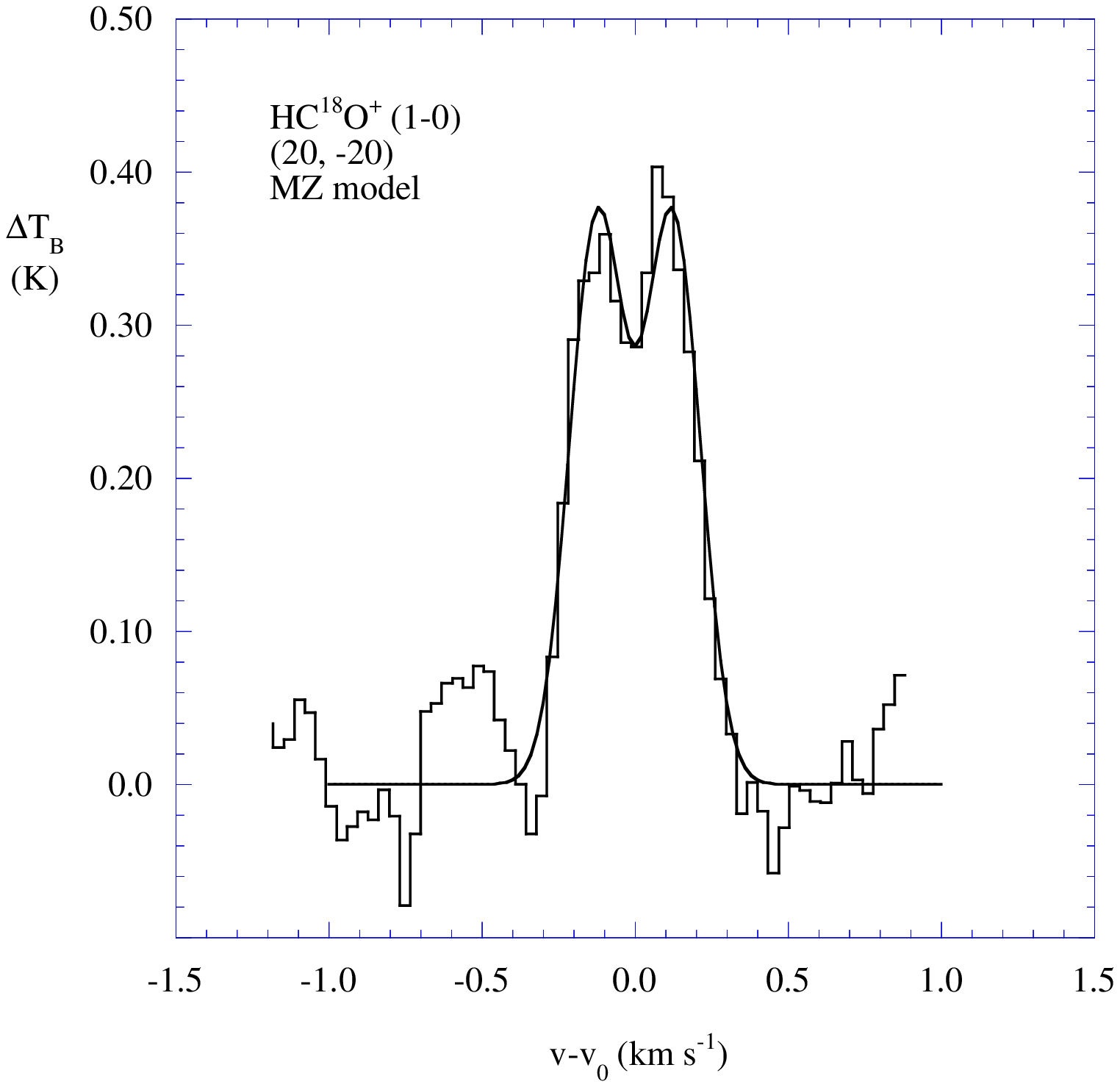}
Fig.9

%\begin{figure}
%\plotone{sgi9259.eps}
%\caption{This is the first figure and it uses sgi9259.eps as
%its EPS figure file. \label{fig1}}
%\end{figure}

% The \plotone and \plottwo commands scale the plot(s) in both dimensions
% so that the horizontal dimension fits in the body of the text.  The
% \plotfiddle command will override any automatic scaling, but often
% requires additional "fiddling" to get the plot to fit on the page.
% The \epsscale command allows the author to simply change the scaling
% of the plot in place, without the additional "fiddling" required by
% \plotfiddle.

\end{document}